%% file: main.tex
\documentclass[a4paper,11pt]{article}
\pdfoutput=1
\usepackage{jheppub}
\usepackage[T1]{fontenc}
\usepackage{caption}
\usepackage{subcaption}
\usepackage{varwidth}
\usepackage{float}
\usepackage{amsmath}
\usepackage{wrapfig}
\usepackage{amssymb}
\usepackage{framed}
\usepackage{indentfirst}
\usepackage{appendix}
\usepackage{xcolor}

\hypersetup{
	hidelinks
}

\title{\boldmath Inflation in the Scale Symmetric Standard Model and Weyl geometry}

\author{Z. Lalak}
\author{and P. Michalak}
\affiliation{Institute of Theoretical Physics, Faculty of Physics,
\\University of Warsaw, ul. Pasteura 5, 02-093 Warsaw, Poland}

\emailAdd{z.lalak@uw.edu.pl}
\emailAdd{pmichalak@fuw.edu.pl}

\abstract{
This work explores the possibility of inflation in a scale-symmetric extension of the Standard Model Higgs sector, where the Higgs field $\phi_1$ is coupled to a singlet scalar, the dilaton $\phi_0$. The two-scalar theory is formulated within Weyl geometry, which modifies the Einstein frame form of the resulting single-field inflationary potential. We extend the analysis to include quantum corrections, incorporating curvature effects in the one-loop effective potential. We find that the resulting spectral index $n_s$ and tensor-to-scalar ratio $r_{0.002}$ can be consistent with the observational constraints. The predicted value $r_{0.002} \lesssim 10^{-3}$ is sufficient to yield a detectable gravitational wave signal. We find the unitarity cutoff in the large-field background, $\Lambda_{UV}\sim M_P/\sqrt{\xi_1}$, which lies below the energy scales relevant during inflation.
}

\begin{document} 
\maketitle
\flushbottom

\input{Introduction}

\input{SSHiggsSS}

\input{Tree_Level}

\input{Quantum_Inflation}

\input{Unitarity}

\input{Conclusions}

\input{Appendices}

\bibliographystyle{JHEP}
\bibliography{Bibliograhy}

\end{document}

%% file: Introduction.tex
\section{Introduction}

This work is motivated by two interconnected directions in theoretical physics: scale symmetry and Weyl geometry. These approaches provide promising frameworks for addressing long-standing questions in cosmology and particle physics. Scale symmetry$-$ the invariance of physical laws under rescalings of length and energy$-$provides a natural path for understanding the origin of mass scales in the Universe. When combined with Weyl geometry \cite{1, 2, 3, Ghilencea:2019rqj, ODA1, ODA3, ODA4, ODA5, Ghilencea:2020rxc, Ghilencea:2021lpa, Ghilencea:2021jjl, Ghilencea:2022lcl}, an extension of Riemannian geometry that preserves scale invariance at the level of action, this framework becomes even more compelling. 

The integration of scale symmetry with Weyl geometry offers several advantages. Firstly, it provides a natural setting for the dynamical generation of fundamental mass scales, including the Higgs mass $m_H$, the Higgs vacuum expectation value $v_h$, and the Planck mass $M_P$, without introducing explicit mass terms in the Lagrangian. This is achieved through a~single scalar field, the dilaton, which can act as a dynamical scale-setting field. Its geometric origin can be associated with a Weyl curvature-squared term, $\Tilde{R}^2$ \cite{1, 2, 3, Ghilencea:2019rqj, ODA1, ODA3, ODA4, ODA5, Ghilencea:2020rxc, Ghilencea:2021lpa, Ghilencea:2021jjl, Ghilencea:2022lcl}. Consequently, mass scales in such a framework emerge from the spontaneous breaking of the scale symmetry rather than being fixed parameters of the theory \cite{SZAP1, SZAP2, SZAP3, ROSS1, ROSS2, ROSS3, Salvio:2014soa, Karananas:2021gco, Lalak:2022wyu}.

In addition, this framework provides a natural setting for exploring inflationary cosmology. Inflation$-$a period of rapid exponential expansion in the early Universe$-$is crucial for explaining the large-scale homogeneity and isotropy of the Universe. The scale-symmetric Higgs-dilaton model embedded in Weyl geometry offers a mechanism for slow-roll inflation that can satisfy current observational constraints. 

Single-field Higgs inflation scenarios have been extensively studied in the literature (see, e.g., \cite{Bezrukov:2007ep, Barbon:2009ya, DeSimone:2008ei, Bezrukov:2010jz, Ghoshal:2024ycp}). In this paper, we explore the inflationary dynamics of the scale-symmetric Standard Model extended by a scalar dilaton field within the context of Weyl conformal geometry. Tree-level inflationary scenarios in similar Higgs-dilaton models have been extensively studied in the literature \cite{Lebedev:2011aq, Lee:2018esk, SZAP2, Ferreira:2019zzx, ROSS3, Enqvist:2014zqa, Piani:2022gon, Karananas:2023zgg}, including their formulation within Weyl geometry \cite{Ghilencea:2020rxc, Ghilencea:2019rqj, 3} and the influence of quantum corrections \cite{Lebedev:2021xey, Lee:2013nv, Lerner:2011ge, SZAP3}. The novelty of this work lies in three key aspects:  
\begin{enumerate}
    \item The explicit inclusion of the dilaton kinetic term in the Weyl geometric framework, crucial for dynamical mass generation;
    \item The incorporation of quantum corrections arising in the Einstein frame Lagrangian due to the coupling to Weyl geometry, which significantly affects the shape of the inflationary potential;
    \item The inclusion of propagator suppression factors \cite{Lerner:2009xg, Lee:2013nv, Salopek:1988qh, Lerner:2011ge}, which modify commutation relation for field non-minimally coupled to gravity with non-canonical kinetic terms in the Einstein frame.
\end{enumerate}

The paper is organized as follows. In Section \ref{Section:Scale_Sym_Higgs_Sector}, we introduce the Higgs-dilaton model in Weyl geometry, constructing a Jordan frame Lagrangian invariant under Weyl conformal transformations and deriving the corresponding Einstein frame theory. Section \ref{Section:Tree_Level} analyzes the classical inflationary dynamics in the slow-roll regime, while Section \ref{Section:Quantum_Inflation} incorporates quantum corrections. Both sections identify viable parameter space regions consistent with current experimental constraints, including predictions for the gravitational wave spectrum. Section \ref{Section:Unitarity} discussed potential unitarity violation in the $W^{\pm}$ boson scattering amplitude. 

Appendix \ref{Appendix:Weyl_geometry} provides a concise overview of Weyl geometry. Appendix \ref{Appendix:Parameter_space} summarizes parameter space constraints resulting from the tree-level Jordan frame potential and experimental bounds. Appendix \ref{Appendix:Beta_functions} details the renormalization procedure, including the beta functions with suppression factors.

%% file: SSHiggsSS.tex
\section{Scale-Symmetric Higgs Scalar Sector}
\label{Section:Scale_Sym_Higgs_Sector}

In this paper, we explore a scale-symmetric extension of the Standard Model Higgs scalar sector, incorporating an additional singlet scalar field, the dilaton $\phi_0$, which emerges naturally from Weyl geometry (see Appendix \ref{Appendix:Weyl_geometry}). The Higgs doublet of $SU(2)_L$ group is parametrized as:
\begin{equation}
    \Phi = \frac{1}{\sqrt{2}}\begin{pmatrix} G_1+i G_2 \\ \phi_1+v+iG_3\end{pmatrix},
\end{equation}
where $\phi_1$ is the neutral Higgs component with vacuum expectation value $v$, and $G_i$ are the Goldstone bosons associated with the broken generators of $SU(2)_L\times U(1)_Y$ symmetry group. We focus on the scalar sector of the theory, which remains invariant under Weyl conformal transformations (\ref{ConfTrans}). In this framework, both the Higgs field $\phi_1$ and the dilaton $\phi_0$ are non-minimally coupled to Weyl gravity and, consequently, to the Weyl vector field $\omega_{\mu}$. The Lagrangian for this system takes the form:
\begin{equation}
    \frac{\mathcal{L}}{\sqrt{g}} = -\frac{1}{12}\Big(\xi_0\phi_0^2+\xi_1\phi_1^2\Big)\Tilde{R}-\frac{1}{4}\Tilde{F}_{\mu\nu}\Tilde{F}^{\mu\nu}+\frac{1}{2}\Tilde{D}_{\mu}\phi_0\Tilde{D}^{\mu}\phi_0+\frac{1}{2}\Tilde{D}_{\mu}\phi_1\Tilde{D}^{\mu}\phi_1-V(\phi_0,\phi_1).
    \label{LWplusHiggs}
\end{equation}
Here $\Tilde{R}$ denotes the Weyl curvature scalar (\ref{Tilde_R_def}), $\Tilde{F}_{\mu\nu}\Tilde{F}^{\mu\nu}$ represents the kinetic term of the Weyl vector field (\ref{Fmunu_Weyl}), and the covariant derivative for scalars $\Tilde{D}_{\mu}\phi_i$ is given by equation (\ref{Tilde_Dmu}). A key aspect of this formulation is the inclusion of the kinetic term for the dilaton field, as this field plays a fundamental role in generating mass scales. In particular, it leads to the emergence of the renormalization scale $\mu(\langle\phi_0\rangle)$\footnote{A scale-symmetric formulation of dimensional regularization can be considered with the conventional constant $\mu$ replaced by a dilaton-dependent function, $\mu(\phi_0)$. This approach ensures scale-invariant quantum corrections, as previously studied in \cite{SZAP4, LOG, GHILENCEA, Olszewski:2019elh, TAMARIT, Ghilencea:2017yqv, Lalak:2018bow}. Once scale symmetry is spontaneously broken$-$when the dilaton field acquires its vev $\langle\phi_0\rangle-$the standard subtraction scale $\mu(\langle\phi_0\rangle)$ emerges, along with other fundamental mass scales such as the Higgs mass $m_H$ and Planck mass $M_P$ (see Appendix \ref{Appendix:Parameter_space} for further discussion).}. The scale symmetric potential $V(\phi_0,\phi_1)$ is given in (\ref{Vtreelevel}).

\begin{figure}[!t]
\vspace{-10pt}
     \centering
     \begin{subfigure}[c]{0.48\textwidth}
         \centering
         \includegraphics[width=\textwidth]{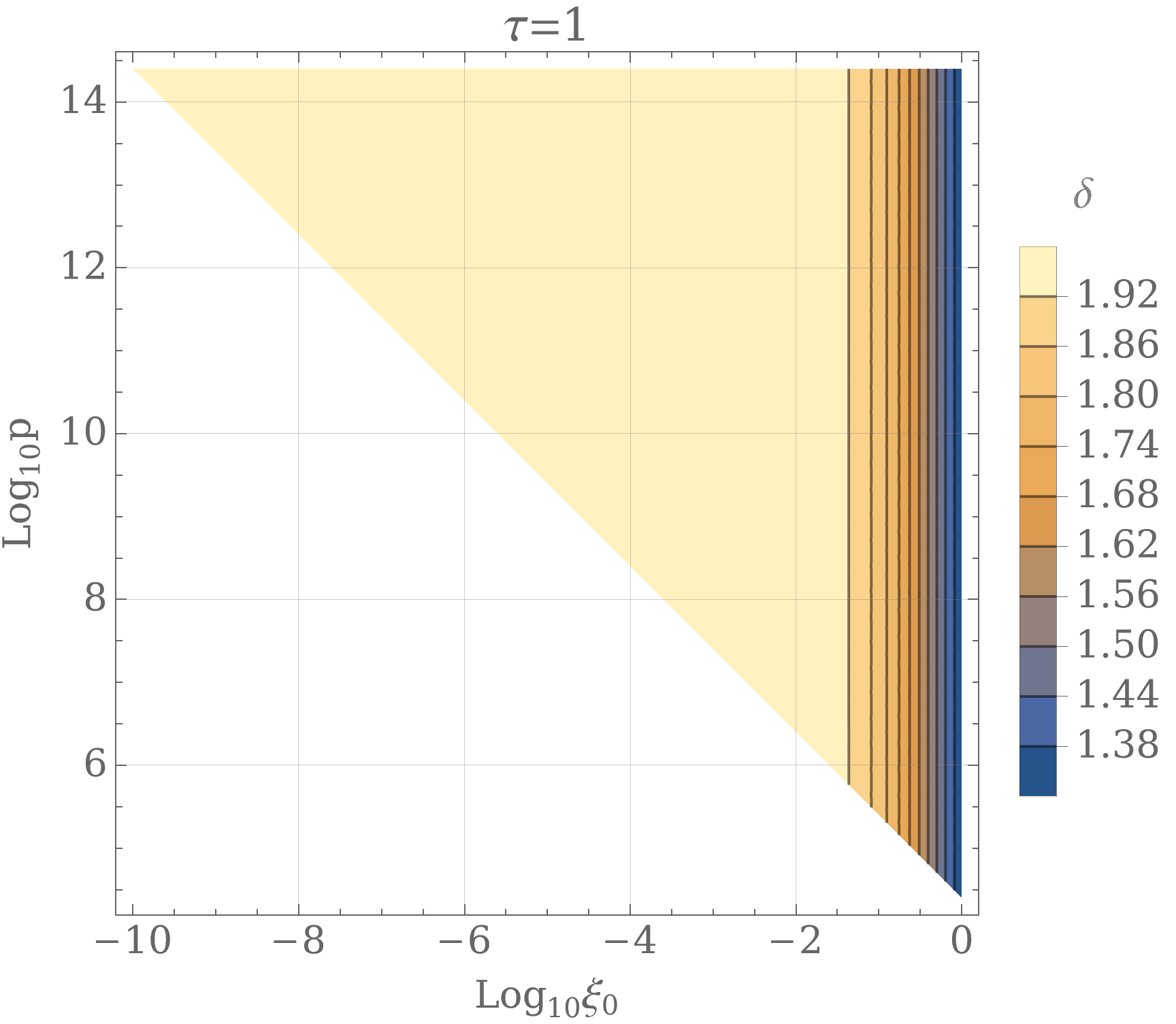}
     \end{subfigure}
     \hfill
     \begin{subfigure}[c]{0.48\textwidth}
         \centering
         \includegraphics[width=\textwidth]{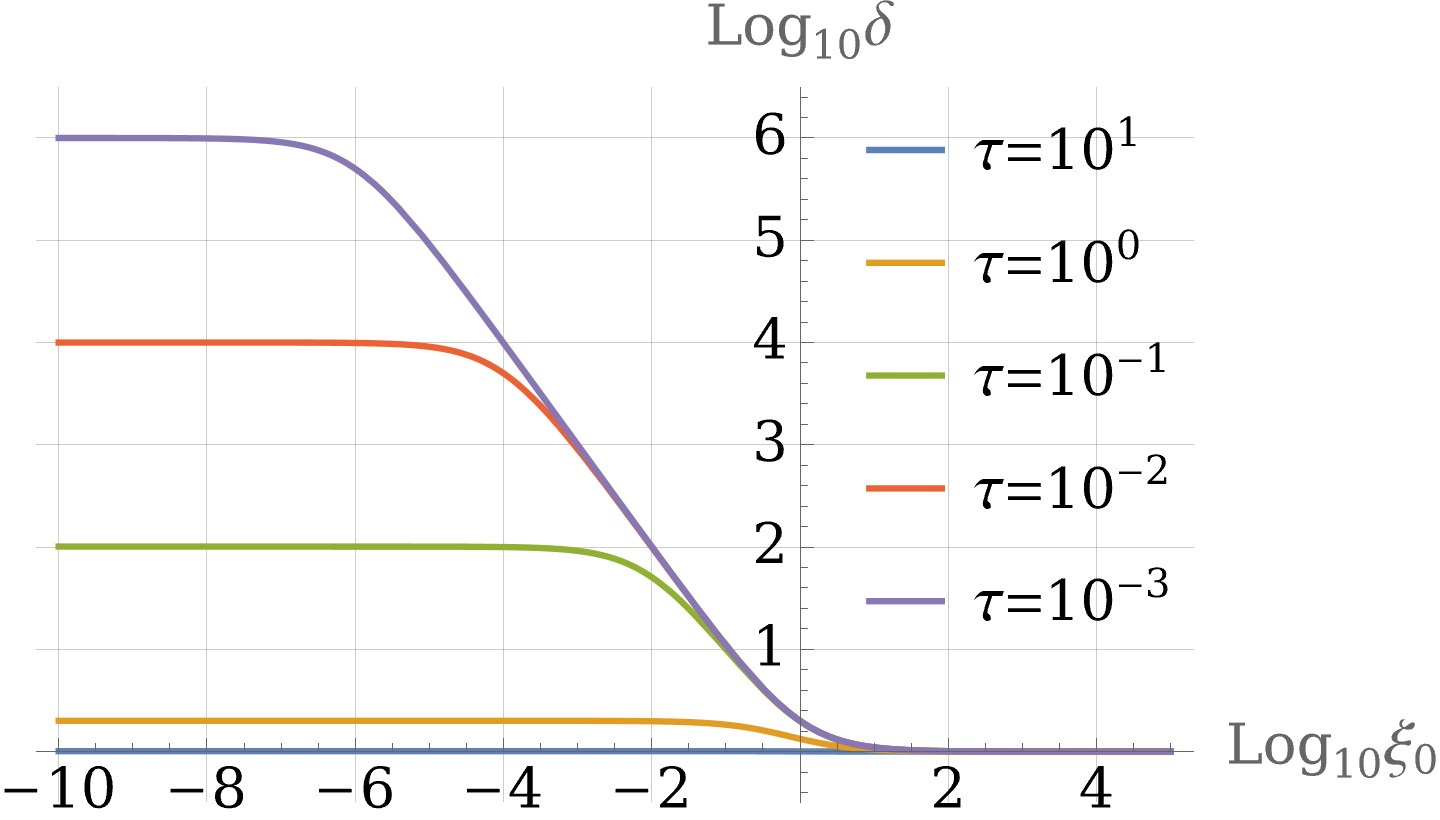}
    \end{subfigure}
    \vspace{-5pt}
    \caption{Plots of $\delta $ for different values of $\xi_i$ and $\tau$, where $\xi_1=p\cdot\xi_0$. The exclusion of certain regions in the $\xi_i$ parameter space follows from subsequent argumentation, as illustrated in Figure \ref{Vmax_fulfilled}. The right plot is evaluated for $p=10^{14.4}$.}
    \label{mw2_tau}
    \vspace{-8pt}
\end{figure}

\begin{figure}[!t]
     \centering
     \begin{subfigure}[c]{0.48\textwidth}
         \centering
         \includegraphics[width=\textwidth]{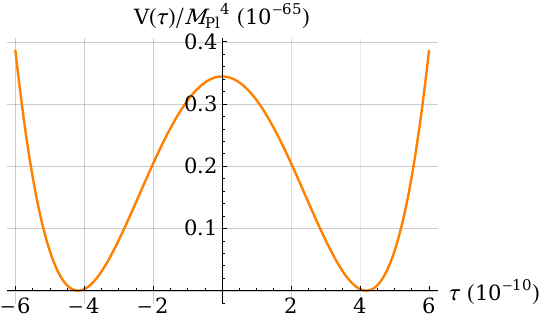}
     \end{subfigure}
     \hfill
     \begin{subfigure}[c]{0.48\textwidth}
         \centering
         \includegraphics[width=\textwidth]{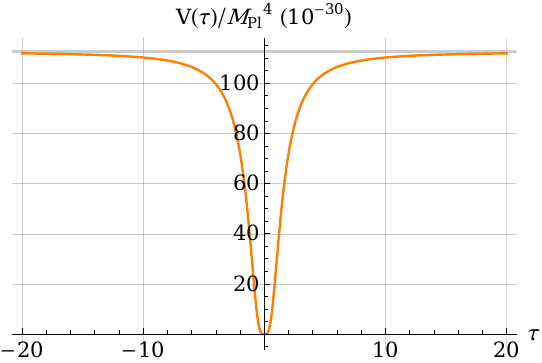}
    \end{subfigure}
    \caption{Plots of the potential $V(\tau)$ from equation (\ref{Vtau_infl}) for $\xi_0=10^{15}$ and $\xi_1=10^{10}$. For small $\tau$, two degenerate non-zero minima are visible, while for large $\tau$, the potential becomes nearly flat, allowing for the possibility of slow-roll inflation.}
    \label{V_infl_plots}
\end{figure}

Expanding $\Tilde{R}$ and $\Tilde{D}_{\mu}\phi_i$ according to equations (\ref{Tilde_R_def}) and (\ref{Tilde_Dmu}), and neglecting total derivative terms, we obtain:
\begin{equation}
    \frac{\mathcal{L}}{\sqrt{g}} = -\frac{1}{12}\rho^2 R-\frac{q}{4}K_{\mu}\omega^{\mu}+\frac{q^2}{8}K\omega_{\mu}\omega^{\mu}+\frac{1}{2}\partial_{\mu}\phi_0\partial^{\mu}\phi_0+\frac{1}{2}\partial_{\mu}\phi_1\partial^{\mu}\phi_1-\frac{1}{4}\Tilde{F}_{\mu\nu}\Tilde{F}^{\mu\nu} - V(\phi_0,\phi_1),
\end{equation}
where we defined:
\begin{equation}
    \rho^2 = \xi_0\phi_0^2+\xi_1\phi_1^2,
\end{equation}
\begin{equation}
    K = \rho^2 + \phi_0^2 +\phi_1^2 = (\xi_0+1)\phi_0^2+(\xi_1+1)\phi_1^2, \qquad K_{\mu} = \partial_{\mu} K.
\end{equation}
To transform the theory into the Einstein-frame, we apply the conformal transformation (\ref{ConfTrans}) to the metric and the Riemannian Ricci scalar $R$:
\begin{equation}
    g_{\mu\nu} \quad \rightarrow \quad \hat{g}_{\mu\nu} = \Omega^2 g_{\mu\nu},
\end{equation}
\begin{equation}
    -\frac{1}{12}\rho^2 R\quad \rightarrow \quad -\frac{1}{2}M^2\hat{R}+\frac{3M^2}{4}\hat{g}^{\mu\nu}\big(\partial_{\mu}\ln\Omega^2\big)\big(\partial_{\nu}\ln\Omega^2\big),
\end{equation}
with:
\begin{equation}
    \Omega^2 = \frac{\rho^2}{6M^2} = \frac{\xi_0\phi_0^2+\xi_1\phi_1^2}{6M^2}.
    \label{Omega_definition}
\end{equation}
The resulting Lagrangian is:
\begin{equation}
\begin{split}
    \frac{\hat{\mathcal{L}}}{\sqrt{\hat{g}}}  = & -\frac{1}{2}M^2 \hat{R} +\frac{1}{\Omega^2}\Big[\frac{1}{2}\partial_{\mu}\phi_0\partial^{\mu}\phi_0+\frac{1}{2}\partial_{\mu}\phi_1\partial^{\mu}\phi_1-\frac{q}{4}K_{\mu}\omega^{\mu}+\frac{q^2}{8}K\omega_{\mu}\omega^{\mu}\Big] + \\
     & +\frac{3M^2}{4}\hat{g}^{\mu\nu}\big(\partial_{\mu}\ln\Omega^2\big)\big(\partial_{\nu}\ln\Omega^2\big) -\frac{1}{4}\Tilde{F}_{\mu\nu}\Tilde{F}^{\mu\nu} - \frac{V(\phi_0,\phi_1)}{\Omega^4}.
\end{split}
\label{Lhat}
\end{equation}
The mixing term $K_{\mu}\omega_{\mu}$ can be removed by redefining the Weyl vector field as:
\begin{equation}
    \hat{\omega}_{\mu} = \omega_{\mu}-\frac{1}{q}\partial_{\mu}\ln K,
\end{equation} 
such that the $\omega_{\mu}$ field effectively "eats" the radial component $K$. Omitting the hats, we get:
\begin{equation}
    \begin{split}
    \frac{\mathcal{L}}{\sqrt{g}}  = & -\frac{1}{2}M^2 R +\frac{1}{\Omega^2}\Big[\frac{1}{2}\partial_{\mu}\phi_0\partial^{\mu}\phi_0+\frac{1}{2}\partial_{\mu}\phi_1\partial^{\mu}\phi_1-\frac{1}{8}\frac{K_{\mu}K^{\mu}}{K}+\frac{q^2}{8}K\omega_{\mu}\omega^{\mu}\Big] + \\
     & +\frac{3M^2}{4}g^{\mu\nu}\big(\partial_{\mu}\ln\Omega^2\big)\big(\partial_{\nu}\ln\Omega^2\big) -\frac{1}{4}\Tilde{F}_{\mu\nu}\Tilde{F}^{\mu\nu} - \frac{V(\phi_0,\phi_1)}{\Omega^4}.
\end{split}
\label{Lagr_Einst_frm}
\end{equation}
The kinetic part of the above Lagrangian can be expressed in terms of the scalar fields as:
\begin{equation}
    \frac{\mathcal{L}}{\sqrt{g}}\Big|_{k.t.} =\frac{3 M^2 \left(\xi _0 \left(\xi _0+1\right) \phi _0^2+\xi _1 \left(\xi _1+1\right) \phi _1^2\right) }{\left(\xi _0 \phi _0^2+\xi _1 \phi _1^2\right){}^2 \left(\left(\xi _0+1\right) \phi _0^2+\left(\xi _1+1\right) \phi _1^2\right)}\left(\phi _1 \partial_{\mu}\phi _0-\phi _0 \partial_{\mu}\phi _1\right){}^2.
    \label{L_kt_phi_i}
\end{equation}
Changing to polar coordinates:
\begin{equation}
\begin{split}
    \phi_0 = & \frac{1}{\sqrt{1+\xi_0}}\sqrt{K} \cos \theta,\\
    \phi_1 = & \frac{1}{\sqrt{1+\xi_1}}\sqrt{K} \sin \theta,
\end{split}
\label{polar_coordinates}
\end{equation}
and denoting $\tau = \tan\theta$, we obtain the following Lagrangian:
\begin{equation}
    \frac{\mathcal{L}}{\sqrt{g}} = -\frac{1}{2}M^2R-\frac{1}{4}\Tilde{F}_{\mu\nu}\Tilde{F}^{\mu\nu}+\frac{q^2}{8}\frac{K}{\Omega^2}\omega_{\mu}\omega^{\mu}+\frac{1}{2}F^2(\tau)\big(\partial_{\mu}\tau\big)\big(\partial^{\mu}\tau\big)-V(\tau),
    \label{L_infl}
\end{equation}
where:
\begin{equation}
    F^2(\tau) = 6M^2\big(\xi_0+1\big)\big(\xi_1+1\big)\frac{\xi_0+\xi_1\tau^2}{\big(\xi_0(\xi_1+1)+\xi_1\tau^2(\xi_0+1)\big)^2\big(\tau^2+1\big)},
    \label{Ftau2}
\end{equation}
\begin{equation}
    V(\tau) = 36M^4\frac{\lambda_0\big(\xi_1+1\big)^2+\lambda_1\big(\xi_0+1\big)\big(\xi_1+1\big)\tau^2+\lambda_2\big(\xi_0+1\big)^2\tau^4}{\big(\xi_0(\xi_1+1)+\xi_1\tau^2(\xi_0+1)\big)^2}.
    \label{Vtau_infl}
\end{equation}
The resulting Lagrangian contains the canonical Einstein-Hilbert term, $-\frac{1}{2}M^2 R$, which identifies the $M$ scale with the Planck mass, $M_P$. The Weyl vector field acquires a mass given by:
\begin{equation}
    m_{\omega}^2(\tau) = \frac{q^2}{4}\frac{K}{\Omega^2} = \frac{3q^2}{2}M^2\frac{(\xi_0+1)(\xi_1+1)(\tau^2+1)}{\xi_0(\xi_1+1)+\xi_1\tau^2(\xi_0+1)} = \frac{3q^2}{2}M^2\cdot \delta.
\end{equation}
For the values of $\xi_i$ in our analysis, the $\delta$ parameter remains larger than 1 and $m_{\omega}^2 \gtrsim \frac{3q^2}{2}M^2$. The Weyl gauge coupling $q$ can have values larger than 1, and $\omega_{\mu}$ mass can be of the order of the Planck mass scale or larger\footnote{One possible way to constrain the value of $q$ is through the Weak Gravity Conjecture in the presence of massless scalar fields. Within this theoretical framework, the gauge coupling $q$ is expected to satisfy $q\lesssim 4$.}. Figure \ref{mw2_tau} further supports this approximation. Consequently, for the energies below $m_{\omega}$ scale, the $\omega_{\mu}$ field effectively decouples from the inflaton, for which the Lagrangian reads:
\begin{equation}
    \frac{\mathcal{L}}{\sqrt{g}} = -\frac{1}{2}M^2R+\frac{1}{2}F^2(\tau)g^{\mu\nu}\partial_{\mu}\tau\partial_{\nu}\tau-V(\tau).
    \label{Lagr_infl_eoms}
\end{equation}
The stationary solutions for $\tau$ are given by:
\begin{equation}
    \langle\tau\rangle = 0,\qquad \langle\tau^2 \rangle= \frac{\left(\xi _1+1\right) \left(2 \lambda _0 \xi _1-\lambda _1 \xi _0\right)}{\left(\xi _0+1\right) \left(2 \lambda _2 \xi _0-\lambda _1 \xi _1\right)} \xrightarrow{\lambda_0=\lambda_1^2/4\lambda_2} -\frac{\lambda_1}{2\lambda_2}\cdot\frac{1+\xi_1}{1+\xi_0}.
    \label{tau_vevs}
\end{equation}
and the potential values at these stationary points are\footnote{We provide a detailed analysis of parameter space constraints derived from the Jordan-frame Lagrangian for the Higgs and dilaton in Appendix \ref{Appendix:Parameter_space}. Defining $\xi_1=p\cdot\xi_0$, the parameter space of the model can be described using two independent variables: $\xi_0$ and $p$.}:
\begin{equation}
    V\big(\langle\tau\rangle=0\big) = 9M^4\frac{\lambda_1^2}{\lambda_2\xi_0^2}>0, \qquad V\big(\langle\tau\rangle\neq 0\big) = 0.
\end{equation}
Figure \ref{V_infl_plots} illustrates the potential (\ref{Vtau_infl}) and its characteristic features for small and large values of $\tau$.

\begin{figure}[!t]
\vspace{-5pt}
    \begin{subfigure}[c]{0.38\textwidth}
         \centering
         \includegraphics[width=\textwidth]{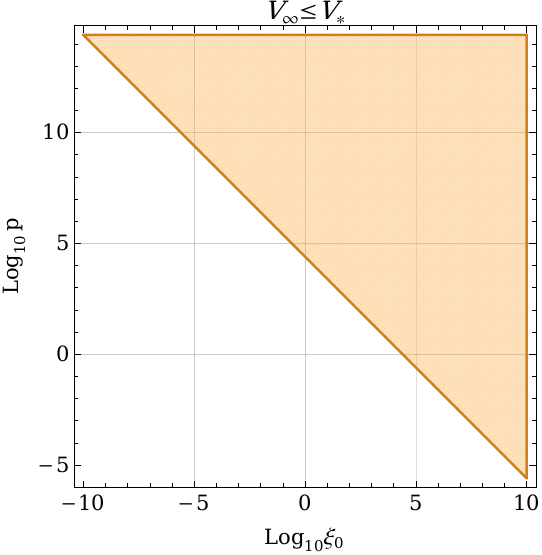}
         \caption{}
         \label{Vmax_fulfilled}
     \end{subfigure}
     \hfill
     \begin{subfigure}[c]{0.59\textwidth}
         \centering
         \includegraphics[width=\textwidth]{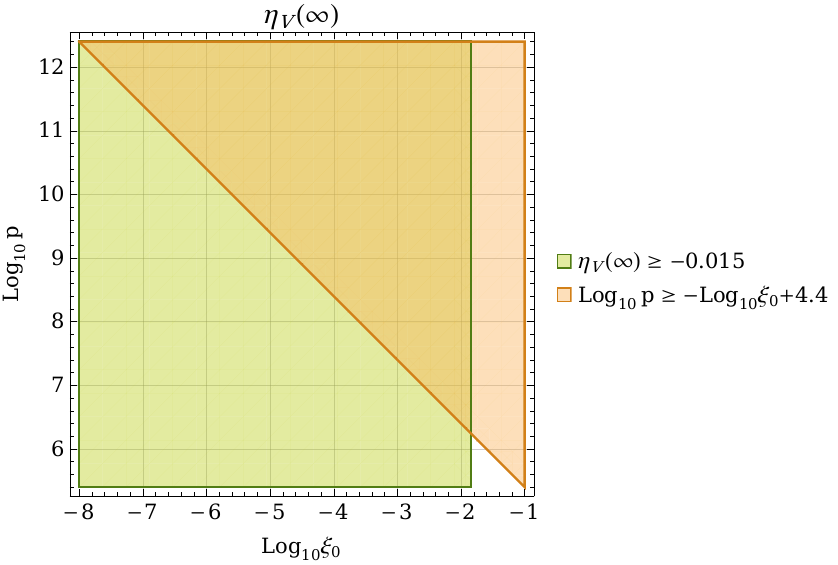}
    \caption{}
    \label{etainfinity}
    \end{subfigure}
    \vspace{-10pt}
    \caption{(a): Region plot of the parameter space for which $V_{\infty}\leq V_*$. The constraint can be approximated by the inequality $\log_{10}p\geq -\log_{10}\xi_0+4.4$. (b): Plot illustrating the constraint $\eta_V(\infty) > -0.0095-0.0030$, resulting from (\ref{etainf_definition}), limiting $\xi_0$ to $\log_{10}\xi_0 \lesssim -1.727$.}
    \vspace{-5pt}
\end{figure}

Special attention should be given to the asymptotic value of the inflaton potential (\ref{Vtau_infl}) for large $\tau$, which approaches a constant:
\begin{equation}
    V_{\infty} = \lim_{\tau\rightarrow\infty}V(\tau) = \frac{36\lambda_2}{\xi_1^2}M_{Pl}^4,
    \label{V_infl_infinity}
\end{equation}
representing the maximum potential energy that could be reached during inflation. The Planck 2018 data bounds this value from above as \cite{Planck:2018jri}:
\begin{equation}
    V^{1/4}_{*} < 1.6\cdot 10^{16} \textrm{GeV}.
    \label{Vbound}
\end{equation}
\noindent This leads to a constraint on the parameter space:
\begin{equation}
    \log_{10}p\geq -\log_{10}\xi_0+4.4,
    \label{p_x0_inequality}
\end{equation}
as illustrated in Figure \ref{Vmax_fulfilled}. The values of non-minimal coupling $\xi_1$ used in this work require a brief discussion. Despite the large value of $\xi_1$, the consistency of the theory is preserved. First, $\xi_1$ sets the unitarity cutoff $\Lambda_{UV}$. However, as demonstrated in Section \ref{Section:Unitarity}, the inflationary energy scale remains safely below $\Lambda_{UV}$. Second, $\xi_1$ does not directly enter the beta functions of the scalar potential or SM couplings. Its running is dominated by Yukawa and gauge contributions (\ref{all_beta_functions}), leading to a moderate evolution of $\xi_1$ by up to approximately 160$\%$. Nevertheless, the quantum RG-improved potential derived in Section \ref{Section:Quantum_Inflation} remains close to its tree-level counterpart. Consequently, key inflationary observables, such as the spectral index $n_s$ and tensor-to-scalar ratio $r_{0.002}$, receive only negligible quantum corrections.



%% file: Tree_Level.tex
\section{Slow-Roll Inflation at the Classical Level}
\label{Section:Tree_Level}

\begin{table}[!t]
\vspace{5pt}
\centering
\caption{Constraints on inflationary parameters from ACT DR6 in combination with external datasets \cite{ACT:2025fju, ACT:2025tim}.
The slow-roll parameter $\epsilon_V$ is inferred from the limit on $r_{0.002}$ via $r_{0.002}=16\,\epsilon_V(\tau^*)$ from eq. (\ref{rns_definition}).}
\label{tab:inflation_params}
\begin{tabular}{lcc}
\hline\hline
Parameter & Constraint & Data combination \\
\hline
$n_s$ & $0.974 \pm 0.003$ (68\% CL) & P-ACT + lensing + BAO \\
$r_{0.002}$ & $<0.035$ (95\% CL) & P-ACT + lensing + BAO + BK18 \\
$\eta_V \,[\times 10^{-4}]$ & $-95^{+23}_{-30}$ (68\% CL) & P-ACT + lensing + BAO + BK18 \\
$\epsilon_V$ & $<2.4\times 10^{-3}$ (95\% CL) & inferred from $r_{0.002}$ \\
\hline
\end{tabular}
\vspace{5pt}
\end{table}

The inflaton field $\tau$ has a non-canonical kinetic term in the Lagrangian (\ref{Lagr_infl_eoms}). The corresponding canonical field variable is given by:
\begin{equation}
    \sigma(\tau) = \int_{v_{\tau}}^{\tau}dx F(x).
    \label{canonical_field}
\end{equation}
However, this integral lacks an analytical solution and must be evaluated numerically. To simplify calculations, the slow-roll parameters can be modified to:
\vspace{-5pt}
\begin{equation}
    \epsilon_V(\tau) = \frac{M_{Pl}^2}{2F^2(\tau)}\Bigg(\frac{V'(\tau)}{V(\tau)}\Bigg)^2,
    \label{epsilonV}
\end{equation}
\vspace{-5pt}
\begin{equation}
    \eta_V(\tau) = \frac{M_{Pl}^2}{F^2(\tau)}\Bigg( \frac{V''(\tau)}{V(\tau)} -\frac{F'(\tau)}{F(\tau)}\frac{V'(\tau)}{V(\tau)} \Bigg)
    \label{etaV}
\end{equation}
and the number of e-folds is given by:
\vspace{-5pt}
\begin{equation}
\vspace{-5pt}
    N_e = \int_{\tau^*}^{\tau_{end}}\frac{d\tau}{\sqrt{2\epsilon_V(\tau)}},
    \label{N_e}
\end{equation}
where $\tau_{end}$ marks the end of inflation, defined as $\eta_V(\tau_{end})= -0.0095$\footnote{See Table \ref{tab:inflation_params} for experimental constraints on cosmological parameters}, and $\tau^*$ corresponds the beginning of inflation, satisfying $\epsilon_V(\tau^*)\ll 1$ and $\eta_V(\tau^*)> -0.0095$.

\begin{figure}[!t]
\vspace{-5pt}
    \begin{subfigure}[c]{0.48\textwidth}
         \centering
         \includegraphics[width=\textwidth]{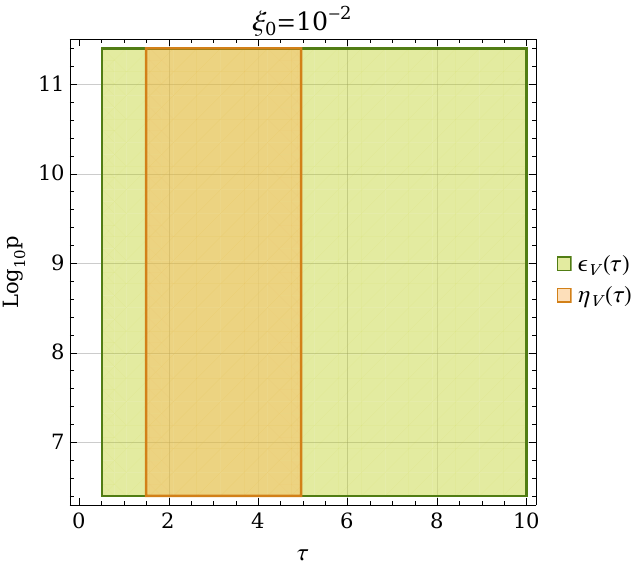}
     \end{subfigure}
     \hfill
     \begin{subfigure}[c]{0.48\textwidth}
         \centering
         \includegraphics[width=\textwidth]{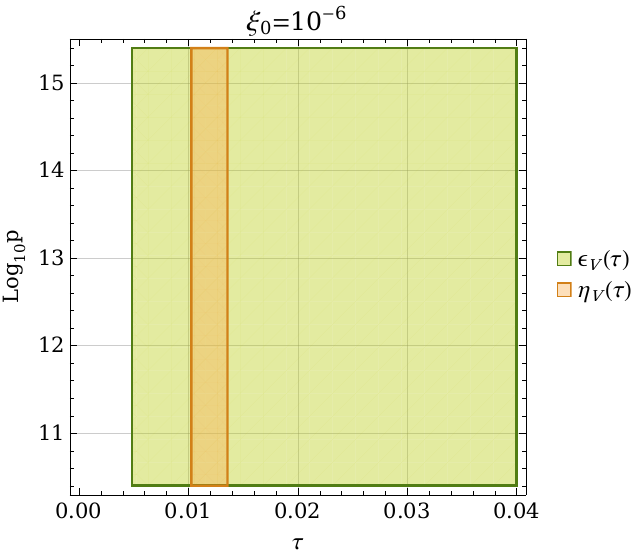}
    \end{subfigure}
    \caption{Plots of the parameter space satisfying the slow-roll conditions from Table \ref{tab:inflation_params}. The minimum $p$ values correspond to the boundary values determined by inequality (\ref{p_x0_inequality}).}
    \label{epsi_eta_regions}
    \vspace{-10pt}
\end{figure}

The second slow-roll parameter $\eta_V(\tau)$ has a non-zero limit when:
\begin{equation}
    \lim_{\tau\rightarrow\infty}\eta_V(\tau) = \frac{1}{3}\Big(-2\xi_0+\frac{\lambda_1\xi_1}{\lambda_2}\Big) = \eta_V(\infty) < 0.
    \label{etainf_definition}
\end{equation}
To fulfill conditions from Table \ref{tab:inflation_params} during inflation, one obtains a constraint $\eta_V(\infty) > -0.0095-0.0030$, illustrated in Figure \ref{etainfinity}. This results in inequality:
\begin{equation}
    \log_{10}\xi_0 \lesssim -1.727.
\end{equation}
The example regions of $\xi_0$, $p$ and $\tau$ where the slow-roll parameters $\epsilon_V(\tau)$ and $\eta_V(\tau)$ fulfill the requirements from Table \ref{tab:inflation_params}, ensuring successful inflation, are illustrated in Figure~\ref{epsi_eta_regions}, where both regions do not depend on the choice of $p$ value. The $\eta_V$ condition imposes a more rigorous constraint on the parameter space. 

\begin{figure}[!t]
     \centering
     \begin{subfigure}[c]{0.48\textwidth}
         \centering
         \includegraphics[width=\textwidth]{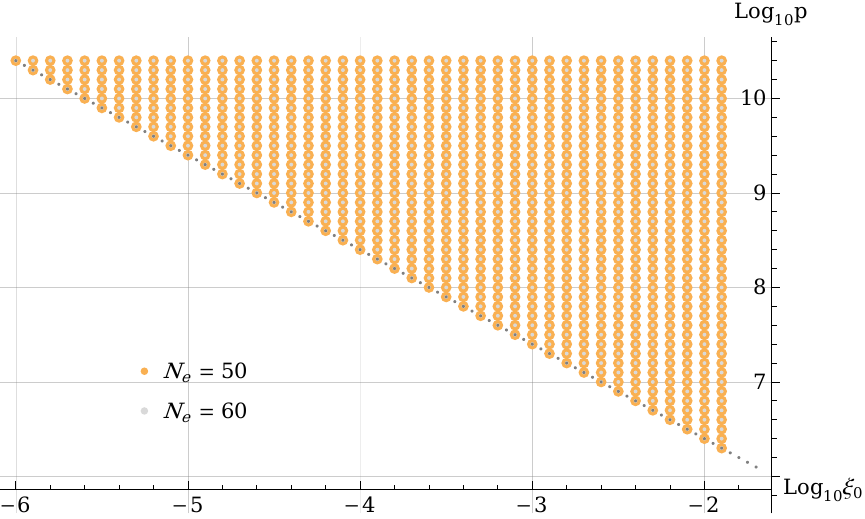}
    \end{subfigure} 
    \hfill
     \begin{subfigure}[c]{0.5\textwidth}
         \centering
         \includegraphics[width=\textwidth]{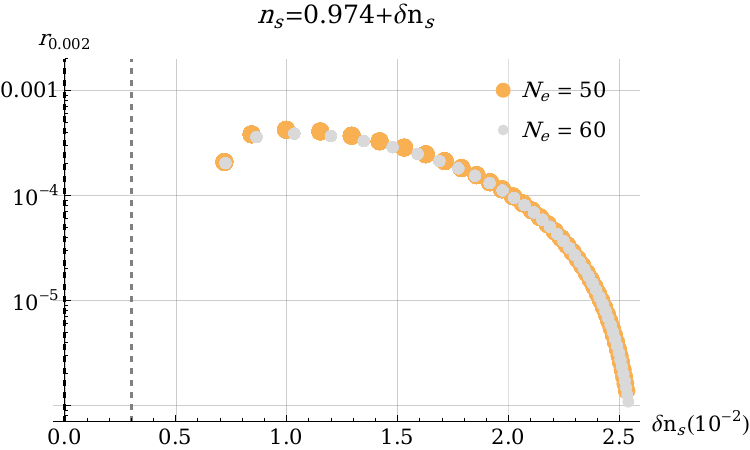}
     \end{subfigure}
    \caption{\textbf{Left:} Parameter space points used to generate the corresponding $r_{0.002}(n_s)$ plot on the right. The dark gray dotted line marks the constraint $\log_{10}p=-\log_{10}\xi_0+4.4$. \textbf{Right:} Tensor-to-scalar ratio $r_{0.002}$ as a function of the scalar spectral index $n_s$ in the model with the inflaton field $\tau$. The black dashed line corresponds to the experimental value, $n_s=0.974$, and the gray dashed line to $n_s=0.974+0.003$. Variations in $\xi_0$ and $p$ lead to small differences in $n_s$, as illustrated in Figure \ref{ns_r_contours}.}
    \label{ns_r_indecies}
\end{figure}

\begin{figure}[!t]
     \centering
     \begin{subfigure}[c]{0.49\textwidth}
         \centering
         \includegraphics[width=\textwidth]{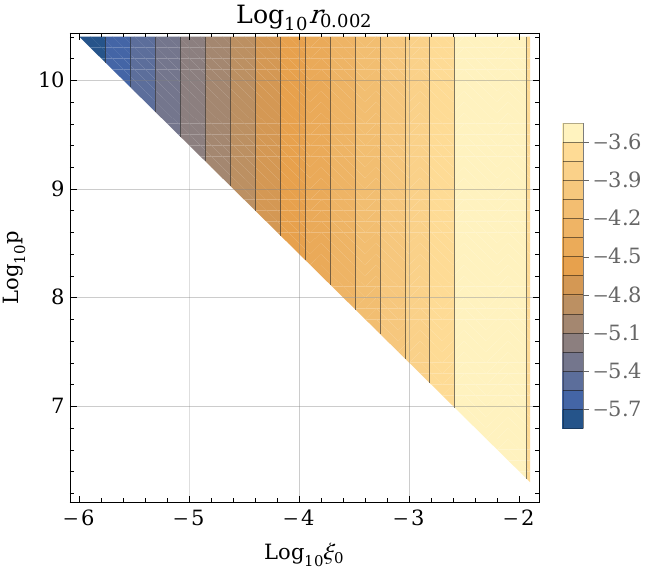}
    \end{subfigure} 
    \hfill
     \begin{subfigure}[c]{0.49\textwidth}
         \centering
         \includegraphics[width=\textwidth]{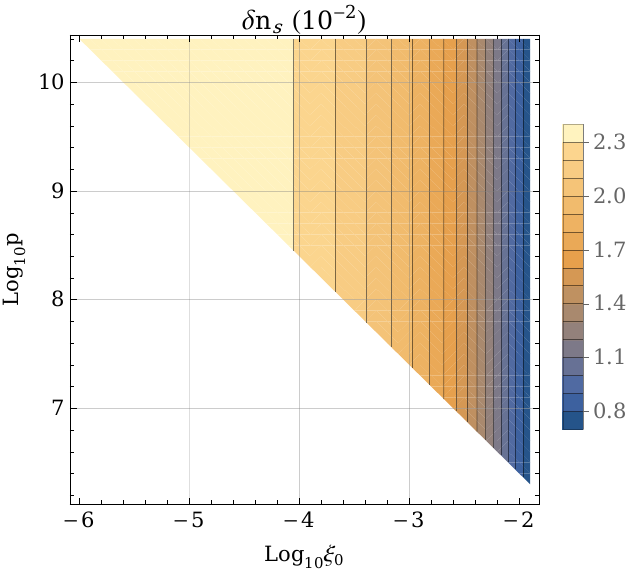}
     \end{subfigure}
     \vspace{-5pt}
    \caption{Plots illustrating the influence of $\xi_0$ and $p$ on $r_{0.002}$ and $n_s$, with the endpoint of inflation set as $\eta_V(\tau_{end}) =-0.0095$, yielding $n_s=0.974+\delta n_s$.}
    \label{ns_r_contours}
    \vspace{-5pt}
\end{figure}

The starting point of inflation, $\tau^*$, is chosen to ensure the number of $e$-folds falls within the range $N_e\in \langle 50,60\rangle$. From this, the inflationary observables are determined using the slow-roll parameters:
\begin{equation}
    n_s = -6\epsilon_V(\tau^*)+2\eta_V(\tau^*)+1,\qquad r = 16\epsilon_V(\tau^*).
    \label{rns_definition}
\end{equation}
Figure \ref{ns_r_indecies} presents the function $r_{0.002}(n_s)$ along with the parameter space region considered in the analysis for $\eta_V(\tau_{end})=-0.0095$. The obtained values of $r_{0.002}$ remain within the experimental bounds (Table \ref{tab:inflation_params}). However, the predicted scalar spectral index $n_s$ exceeds its allowed values. Figure \ref{ns_r_contours} illustrates the impact of $\xi_0$ and $p$ on $r_{0.002}$ and $n_s$. Additionally, the predicted values of $r_{0.002}$ in this model are sufficient to generate a detectable gravitational wave signal, as discussed in Subsection \ref{Subsection:GW_spectrum}. 

\begin{figure}[!t]
     \centering
         \includegraphics[width=0.9\textwidth]{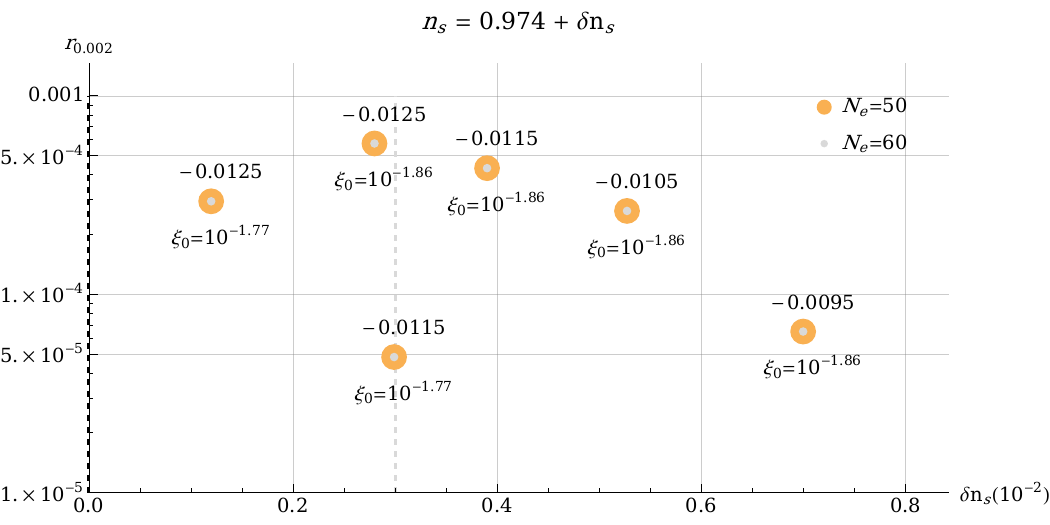}
         \vspace{-5pt}
         \caption{Tensor-to-scalar ratio $r_{0.002}(n_s)$ for $\xi_0=10^{-1.71}$ and $\xi_0=-1.86$ with $p=10^{6.5}$ at different inflationary endpoints, determined by $\eta_V(\tau_{end})$. The numerical labels above the points indicate the corresponding $\eta_V(\tau_{end})$ values. The black dashed line corresponds to the experimental value, $n_s=0.974$, and the gray dashed line to $n_s=0.974+0.003$.}
    \label{rns_classical_diff_etaV}
    \vspace{-5pt}
\end{figure}

\begin{figure}[!t]
     \centering
     \begin{subfigure}[c]{0.49\textwidth}
         \centering
         \includegraphics[width=\textwidth]{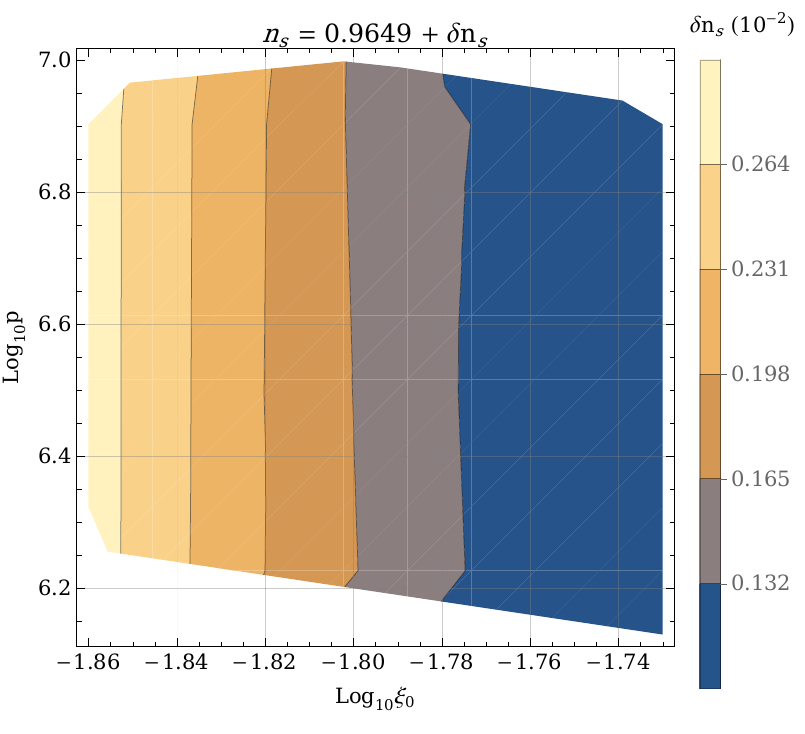}
    \end{subfigure} 
    \hfill
     \begin{subfigure}[c]{0.49\textwidth}
         \centering
         \includegraphics[width=\textwidth]{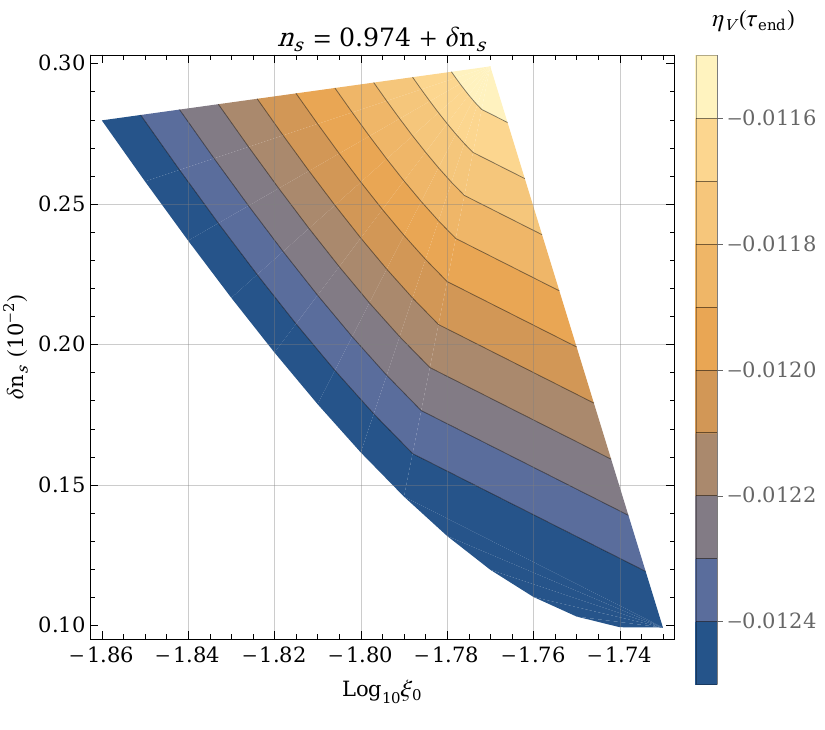}
     \end{subfigure}
    \caption{Plots illustrating the allowed parameter space to obtain $n_s$ consistent with experiments (Table \ref{tab:inflation_params}). In the left plot, the $p$ values were truncated; however, because $n_s$ does not depend on $p$, the constraint applies only to $\xi_0$. For this reason, the right plot highlights exclusively the $\xi_0$ dependence.}
    \label{allowed_class}
\end{figure}

The final value of the scalar spectral index $n_s$ is primarily influenced by the choice of $\eta_V(\tau_{end})$. Therefore, by redefining the end of inflation, $\tau_{end}$, consistently with the observational constraints, it is possible to achieve results in agreement with experimental data. Figure \ref{rns_classical_diff_etaV} presents these revised results for $\xi_0=10^{-1.86}$ and $\xi_0=-1.77$ with $p=10^{6.5}$, chosen to obtain proper $n_s$ value at a given $\eta_V(\tau_{end})$. Figure \ref{allowed_class} further illustrates the parameter space essential to obtain viable cosmological predictions.

In this section, we formulated a scale-symmetric Higgs-dilaton sector within Weyl geometry, where both scalar fields are non-minimally coupled to the Weyl curvature scalar. By performing a conformal transformation to the Einstein frame, we obtained an effective theory in which the Weyl vector field decouples, resulting in a canonical Riemannian gravity sector. At the classical level, the potential for the inflaton field $\tau$ supports a slow-roll inflationary scenario, leading to the tensor-to-scalar ratio $r_{0.002}$ and the scalar spectral index $n_s$ that align with the observational constraints. In the next section, we explore how quantum corrections modify this framework, particularly their influence on the inflationary observables.

%% file: Quantum_Inflation.tex
\section{Slow-Roll Inflation at the Quantum Level}
\label{Section:Quantum_Inflation}

In this section, we extend the analysis of slow-roll inflation within the scale-symmetric Higgs-dilaton model by incorporating quantum corrections to the inflationary potential (\ref{Vtau_infl}). Specifically, we compute the one-loop effective potential and apply renormalization group (RG) improvement. Our analysis is focused on two key aspects:
\begin{enumerate}
    \item \textbf{Choice of the Renormalization Frame}\\
    There is an ongoing discussion regarding whether a gravity-coupled theory should be renormalized in the Jordan or Einstein frame and whether both frames remain equivalent at the quantum level \cite{Falls:2018olk, Kamenshchik:2014waa, Ghilencea:2022xsv, George:2013iia, Ruf:2017xon, SZAP3}. Here, we follow the approach discussed in \cite{Falls:2018olk}, which argues that to obtain a consistent quantum theory regardless of the frame choice, one must account for the changes in the path integral measure when including gravity and its associated metric field. Specifically, finite contributions, absent under a naive conformal transformation of fields, arise when changing frames. Instead of tracking the path integral measure, the issue can be resolved alternatively. Renormalizing in the Einstein frame, one uses the conventional constant renormalization scale $\mu$, while in the Jordan frame, one employs a field-dependent regulator $\mu(\phi)$. The Jordan frame approach aligns with the scale-invariant regularization framework \cite{SZAP4, LOG, GHILENCEA, Olszewski:2019elh, TAMARIT, Ghilencea:2017yqv, Lalak:2018bow}, where $\mu$ depends on the dilaton field. Consequently, the one-loop potential $V_1(\tau)$ for the inflaton field $\tau$ can be derived in the Einstein frame, where the gravity sector is canonical. This allows us to apply standard methods to compute $V_1(\tau)$, as in \cite{Markkanen:2018bfx, Markkanen:2017tun, Czerwinska:2015xwa}. On the other hand, the beta functions for the coupling constants are obtained from the renormalization performed in the Jordan frame with $\mu(\phi_0)$, where their calculation is more straightforward. Since, at the one-loop level, those $\beta$ functions coincide with their $\mu=const$ counterparts \cite{LOG, GHILENCEA, Lalak:2018bow, Olszewski:2019elh, Ghilencea:2016ckm}, their explicit form can be found in \cite{Czerwinska:2015xwa, Lebedev:2021xey}.

    \item \textbf{Propagator Suppression Factors}\\
    Following the approach in \cite{Lerner:2009xg, Lee:2013nv, Salopek:1988qh, Lerner:2011ge}, we account for modifications in the commutation relations that arise when transforming between frames. This introduces the propagator suppression factors $c_{\phi_i}$, which affect the $\beta$ functions obtained in the Jordan frame. These modifications apply to the theories where scalar fields are non-minimally coupled to gravity via $\xi_i\phi_i^2 R$ terms. As described in Section \ref{Section:Tree_Level}, transforming the Jordan frame Lagrangian for the fields $\phi_i$ into the Einstein frame, involves rescaling the metric as $g^{\mu\nu}\rightarrow\Omega^{2}g^{\mu\nu}$. Given the canonical field variable $\sigma$ (\ref{canonical_field}) and the Lagrangian for the inflaton field $\tau$ (\ref{Lagr_infl_eoms}), the canonical commutation relation $\big[\phi(x),\dot{\phi}(y) \big]$ is modified to (for details, see Appendix \ref{Appendix:Beta_functions}):
    \begin{equation}
        \big[\phi_i(x),\dot{\phi}_i(y)\big]  = \frac{i}{a^3}\cdot c_{\phi_i} \cdot \delta^{(3)}\big(\vec{x}-\vec{y}\big),
    \end{equation}
    where the suppression factors $c_{\phi_i}$ are defined as:
    \begin{equation}
        c_{\phi_i} = \frac{1}{\Omega^2\Big(\frac{\partial\sigma}{\partial\phi_i}\Big)^2} = \frac{1}{\Omega^2 F^2(\tau)}\Bigg(\frac{\partial\phi_i}{\partial\tau}\Bigg)^2.
    \label{c_factors_definition}
    \end{equation}
    The resulting $\beta$ functions, along with a comparison of their behavior with and without the inclusion of $c_{\phi_i}$, are provided in Appendix \ref{Appendix:Beta_functions}.
    
\end{enumerate}

For a canonical $\sigma$ field with Lagrangian:
\begin{equation}
    \frac{\mathcal{L}}{\sqrt{g}} = -\frac{1}{2}M_P^2R+\frac{1}{2}g^{\mu\nu}\partial_{\mu}\sigma\partial_{\nu}\sigma-V(\sigma),
\end{equation}
the one-loop effective potential takes the form\footnote{We apply the results from \cite{Markkanen:2018bfx} during the inflation, where the space-time has the de Sitter form, with:
\begin{equation}
    R=12H^2, \qquad R_{\mu\nu}R^{\mu\nu} = 36H^4, \qquad R_{\mu\nu\rho\sigma}R^{\mu\nu\rho\sigma}=24H^4.
\end{equation}} \cite{Markkanen:2018bfx}:
\begin{equation}
    V_{1-\textrm{loop}}(\sigma)\Big|_{\sigma} = V(\sigma)+\frac{1}{64\pi^2}\Bigg[n_{\sigma}\mathcal{M}^4(\sigma)\log\frac{\big|\mathcal{M}^2(\sigma)\big|}{c_{\sigma}\mu^2} + n'_{\sigma}\frac{R^2}{144}\log\frac{\big|\mathcal{M}^2(\sigma)\big|}{\mu^2}\Bigg],
    \label{V1sigma}
\end{equation}
where $c_{\sigma}=e^{3/2}$, $n_{\sigma}=1$, $n_{\sigma}'=-\frac{2}{15}$, $\mathcal{M}^2(\sigma) = m^2(\sigma) - \frac{R}{6}$ and:
\begin{equation}
    m^2(\sigma) = \frac{\partial^2 V(\sigma)}{\partial\sigma^2} = m^2(\tau) = \frac{1}{F^2(\tau)}\Bigg(\frac{\partial^2V(\tau)}{\partial\tau^2}-\frac{\partial V(\tau)}{\partial\tau}\frac{\partial F(\tau)/\partial\tau}{F(\tau)}\Bigg),
    \label{msigma_scale}
\end{equation}
where $V(\tau)$ and $F(\tau)$ are given in equations (\ref{Vtau_infl}) and (\ref{Ftau2}) respectively, $V(\sigma)=V(\tau)$ and consequently $V_{1-\textrm{loop}}(\sigma) = V_{1-\textrm{loop}}(\tau)$. The curvature contribution in the scalar part cannot be neglected, as during the inflation $R\approx 12 H_{inf}^2$, with large $H_{inf}^2$, and this term is dominant over $m^2(\sigma)$.

 The $\phi_1$ field represents the Higgs neutral component, hence, it couples to the SM particles, especially $W^{\pm}$, $Z^0$ bosons, and the top quark, which give the most dominant contribution to the effective potential. The Jordan frame masses are given by:
 \begin{equation}
\begin{split}
    &m^2_W(\phi_1) = \frac{g_2^2}{4}\phi_1^2, \qquad m^2_Z(\phi_1) = \frac{g_1^2+g_2^2}{4}\phi_1^2, \qquad m_t^2(\phi_1) = \frac{y_t^2}{2}\phi_1^2, \\
    & \qquad \qquad \qquad \qquad m_G^2(\phi_0,\phi_1) = 2\lambda_1\phi_0^2+4\lambda_2\phi_1^2,
\end{split}
\end{equation}
where $m_G$ is the mass for the Goldstone bosons $G_i$. After the transformation to Einstein frame and changing to polar coordinates (\ref{polar_coordinates}), masses $m_i^2$ take the form:
\begin{equation}
\begin{split}
    &\hat{m}_{W,Z}^2(\tau) = \frac{m_{W,Z}^2\big(\phi_1(\tau)\big)}{\Omega^2\big((\phi_0(\tau),\phi_1(\tau)\big)}, \qquad \hat{m}^2_t(\tau) = \frac{m_t^2\big(\phi_1(\tau)\big)}{\Omega^2\big((\phi_0(\tau),\phi_1(\tau)\big)}, \\
    & \qquad \qquad \qquad \qquad \hat{m}^2_G(\tau) = \frac{m_G^2\big(\phi_0(\tau),\phi_1(\tau)\big)}{\Omega^4\big((\phi_0(\tau),\phi_1(\tau)\big)}.
\end{split}
\label{mass_scales_hat}
\end{equation}
Consequently, the Standard Model particles contribute to the one-loop effective potential as \cite{Markkanen:2018bfx}:
\begin{equation}
    V_{1-\textrm{loop}}(\tau)\Big|_{SM} = \frac{1}{64\pi^2}\sum_{i=W, Z, t}\Bigg[ n_i\mathcal{M}_i^4(\tau)\log\frac{\big|\mathcal{M}_i^2(\tau)\big|}{c_i\mu^2}+n'_i\frac{R^2}{144}\log\frac{\big|\mathcal{M}_i^2(\tau)\big|}{\mu^2}  \Bigg],
    \label{V1SM}
\end{equation}
where:
\begin{equation}
\begin{array}{llll}
   \mathcal{M}^2_W(\tau)  = \hat{m}_W^2(\tau)+\frac{R}{12}, \qquad &c_W = e^{5/6},& \qquad n_W = 6,& \qquad n'_W = -\frac{34}{5}, \vspace{5pt}\\
   \mathcal{M}^2_Z(\tau) = \hat{m}_Z^2(\tau)+\frac{R}{12}, \qquad &c_Z = e^{5/6}, &\qquad n_Z = 3, &\qquad n'_Z = -\frac{17}{5}, \vspace{5pt}\\
   \mathcal{M}^2_t(\tau)  = \hat{m}_t^2(\tau)+\frac{R}{12}, \qquad &c_t = e^{3/2}, &\qquad n_t = -12,& \qquad n'_t = \frac{38}{5}, \vspace{5pt}\\
   \mathcal{M}^2_G(\tau)  = \hat{m}_G^2(\tau)+\frac{R}{6}\big(\xi_1-1\big), \qquad &c_G = e^{3/2},& \qquad n_G = 3, &\qquad n'_G = -\frac{6}{15}.
\end{array}
\label{M_i^2}
\end{equation}
Finally, the full one-loop effective potential for the inflaton field $\tau$ reads:
\begin{equation}
    V_{1-\textrm{loop}}(\tau) = V_{1-\textrm{loop}}(\tau)\Big|_{\sigma}+V_{1-\textrm{loop}}(\tau)\Big|_{SM},
    \label{V1loop_full}
\end{equation}
with the corresponding contributions given in equations (\ref{V1sigma}) and (\ref{V1SM}).


\subsection{RG Improvement and Choice of the Renormalization Scale}

\begin{figure}[!t]
     \centering
     \begin{subfigure}[c]{0.49\textwidth}
         \centering
         \includegraphics[width=\textwidth]{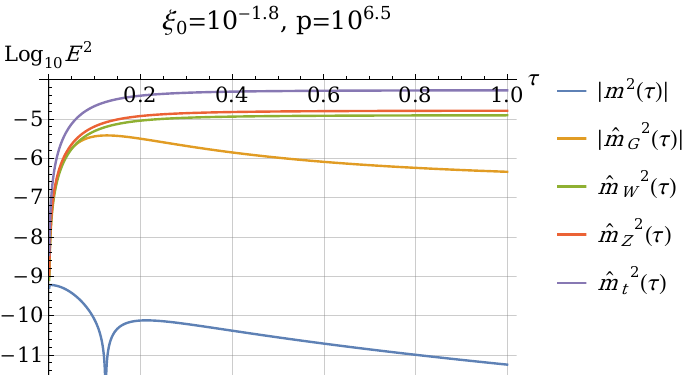}
    \end{subfigure} 
    \hfill
    \begin{subfigure}[c]{0.49\textwidth}
         \centering
         \includegraphics[width=\textwidth]{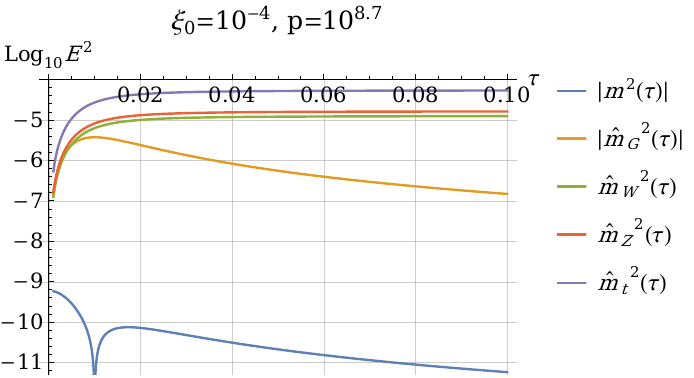}
    \end{subfigure}
    \caption{Plots of the mass scales (\ref{mass_scales_hat}) and $m^2(\tau)$ (\ref{msigma_scale}) for selected parameter points. In the inflationary regime, the dominant energy contributions flatten out.}
    \label{mass_scales}
\end{figure}

A key aspect of the one-loop effective potential analysis is the choice of an appropriate renormalization scale $\mu$ to minimize the logarithmic terms in $V_{1-\textrm{loop}}(\tau)$. Guided by the approximately constant energy scales during inflation, we adopt a choice of a fixed energy scale $\mu$\footnote{As discussed in \cite{Markkanen:2018bfx}, one can choose $\mu$ in a way that the logarithms vanish. However, this prescription introduces an ambiguity in the definition of $\mu$, and the quantum potential behavior can still be compromised. A common alternative is to set $\mu = a\cdot f(\phi) + b\cdot R$, where $f(\phi)$ is a field-dependent function. In the framework considered here, this choice reduces to a constant $\mu$ in the inflationary regime of $\tau$. For this reason, we adopt the constant-$\mu$ approach.}. The tree-level potential $V(\tau)$ becomes nearly flat for at least $\tau\gtrsim\mathcal{O}(10^2)$. The same is true for the mass scales (\ref{M_i^2}), as illustrated in Figure \ref{mass_scales}. The Hubble parameter associated with $V(\tau)$, $H_{inf}$, remains nearly constant throughout slow-roll inflation. Therefore, we choose a constant $\mu$ value and explore two possibilities:
\begin{equation}
    \begin{split}
        & \mu_1 = H_{inf}, \\
        & \mu_2 = M_t\big(\tau=10^2\big),
    \end{split}
\end{equation}
where $\tau=10^2$ is chosen as the value for which the energy scales reach their asymptotic limits and are in the inflationary regime. The Hubble scale $H_{inf}$ is evaluated numerically for each parameter space point so that:
\begin{equation}
    H^2_{inf} = \frac{V_{1-\textrm{loop}}(\tau=10^2,H_{inf},\mu_i)}{3M_P^2}.
\end{equation}
Analysis of the above two $\mu_i$ choices leads to negligible differences in the results. Even though the $\mu_1$ and $\mu_2$ may differ in orders of magnitude, the differences in the couplings values for different $\mu_i$ do not exceed $5\%$. Consequently, we adopt $\mu_1=H_{inf}$ throughout the rest of this work.

We apply the RG improvement by solving the RG equations (\ref{all_beta_functions}) to determine the running of couplings and evaluate their values at $\mu_1$. We promote the inflaton field to $\tau\rightarrow\exp\big(\Gamma(\mu_1)\big)\tau$ as provided in equation (\ref{Gamma_mu}). Plots of the couplings and inflaton field running with $\mu$ energy scale are provided in Appendix \ref{Appendix:Beta_functions}. 

\begin{figure}[!t]
     \centering
     \begin{subfigure}[c]{0.49\textwidth}
         \centering
         \includegraphics[width=\textwidth]{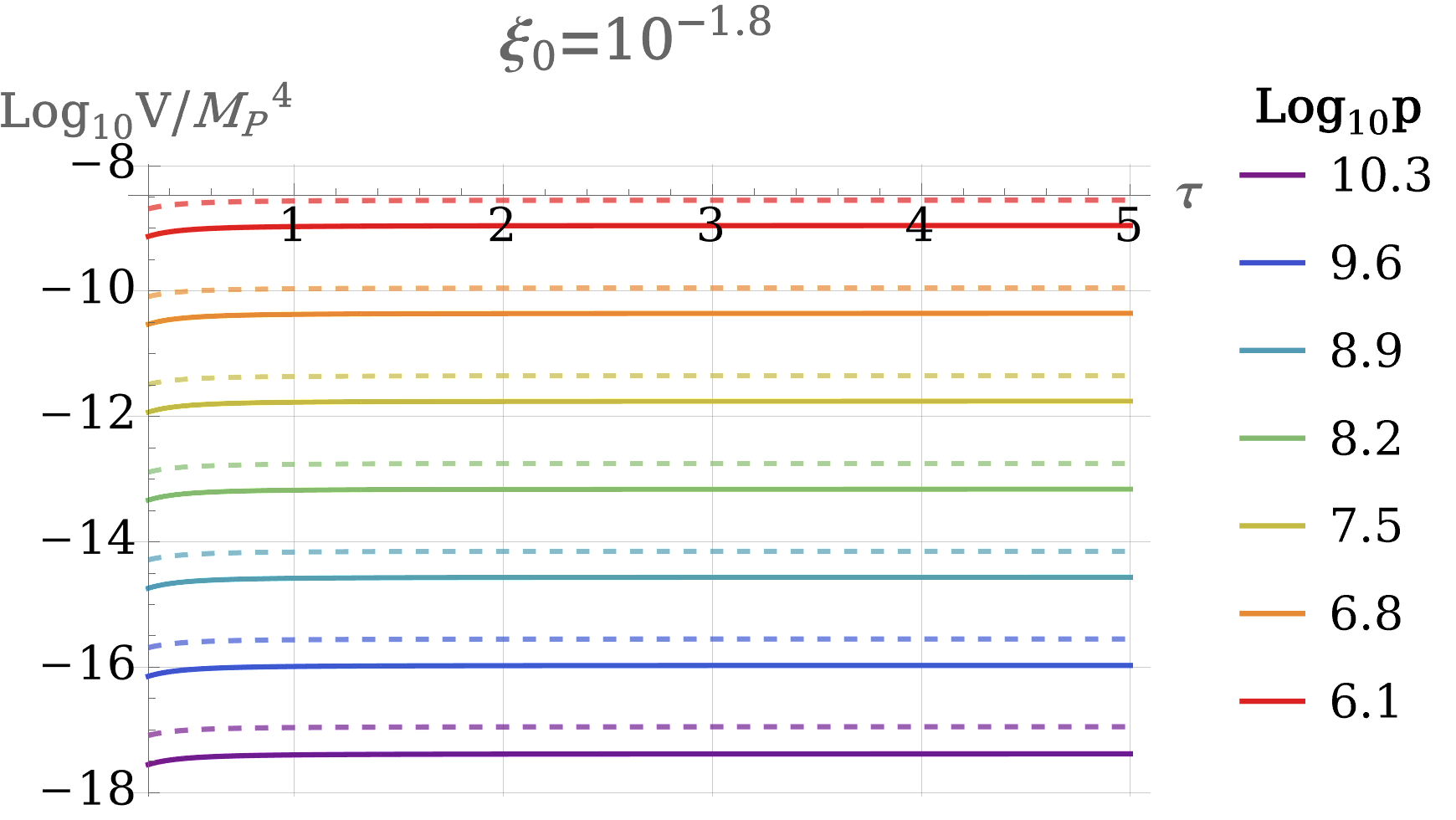}
    \end{subfigure} 
    \hfill
    \begin{subfigure}[c]{0.49\textwidth}
         \centering
         \includegraphics[width=\textwidth]{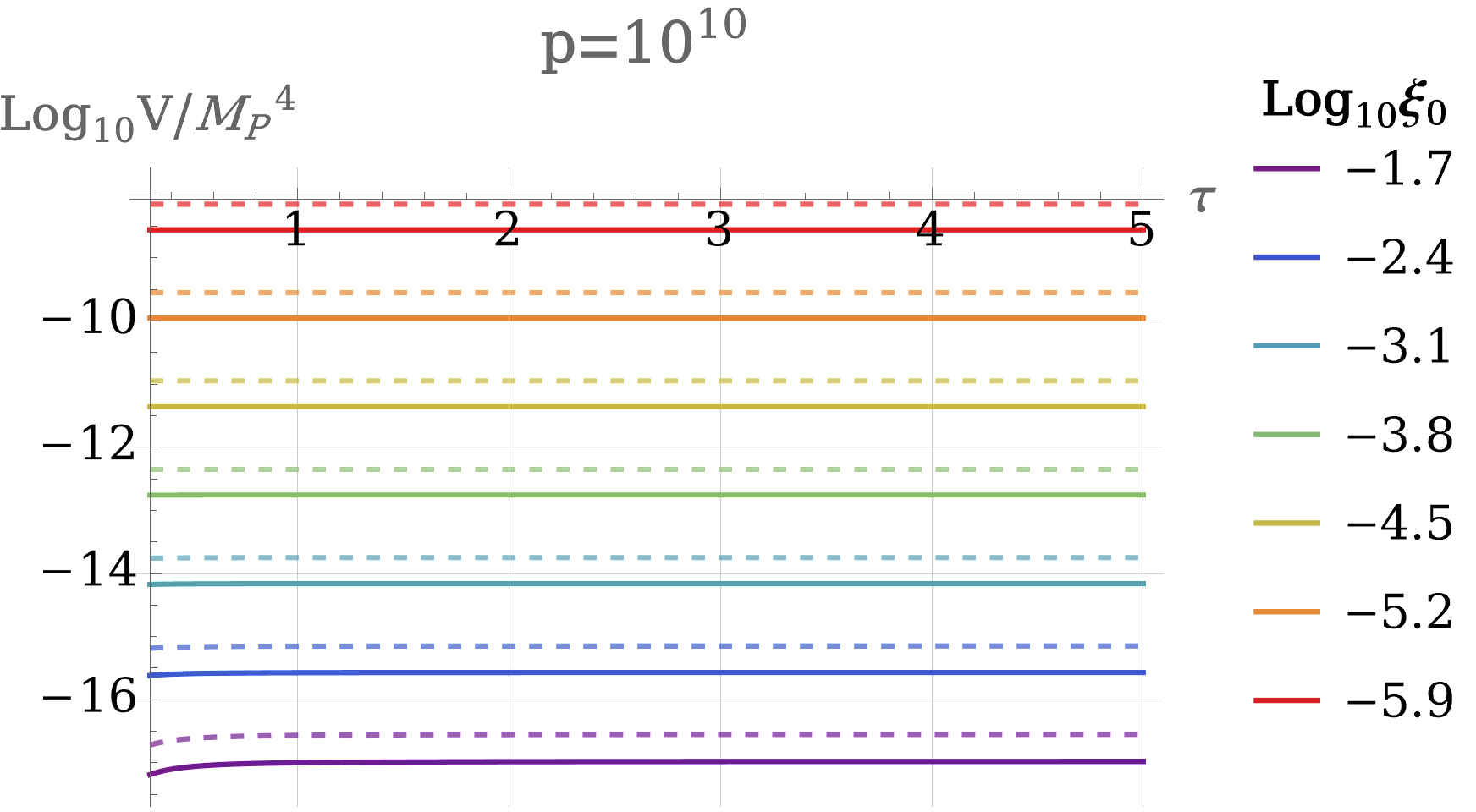}
    \end{subfigure}
    \caption{Plots of the RG-improved effective potential $V_{RGI}(\tau)$ (solid lines) and the tree-level potential $V(\tau)$ (dashed lines) for selected parameter space values. Quantum corrections reduce the potential values slightly, by less than an order of magnitude.}
    \label{VRGI_plots}
\end{figure}

\begin{figure}[!t]
     \centering
     \begin{subfigure}[c]{0.5\textwidth}
         \centering
         \includegraphics[width=\textwidth]{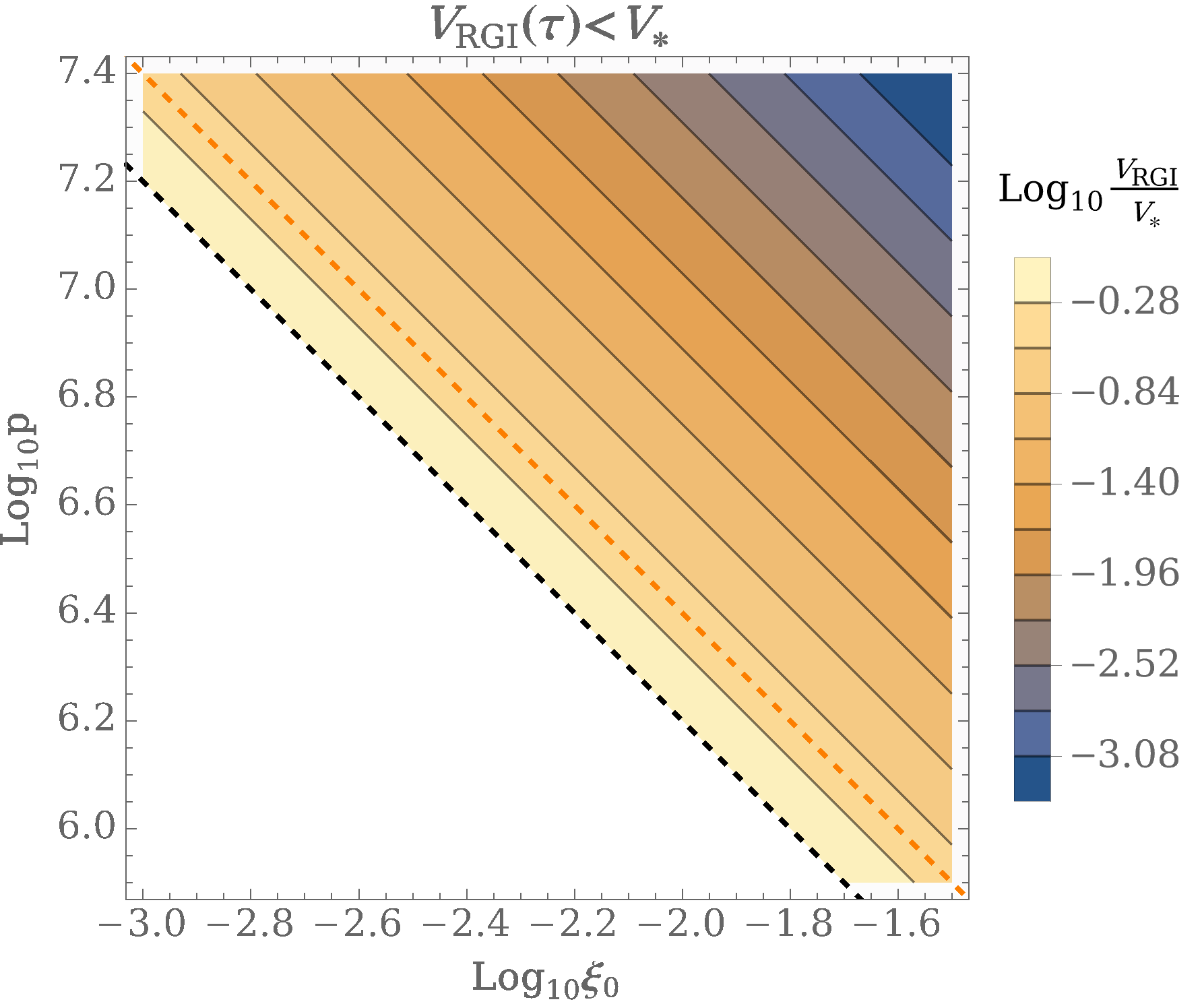}
     \end{subfigure}
    \caption{Region plot of the allowed values of $\xi_0$ and $p$ for which $V_{RGI}(\tau)$ does not exceed the bound (\ref{Vbound}). The result can be approximated by the inequality $\log_{10}p\geq -\log_{10}\xi_0+4.2$, marked by the black dashed line. For comparison, the tree-level result $\log_{10}p\geq -\log_{10}\xi_0+4.4$ is indicated by the orange dashed line.}
    \label{VRGI_regionplt}
\end{figure}

The RG improved potential $V_{1-\textrm{loop}}(\tau,\mu=\mu_1) = V_{RGI}(\tau)$ is presented in Figure \ref{VRGI_plots}, where we compare it to the classical potential $V(\tau)$. The constraints on the parameter space for which $V_{RGI}(\tau)$ does not exceed the bound (\ref{Vbound}) are illustrated in Figure \ref{VRGI_regionplt}. These constraints lead to the approximate inequality $\log_{10}p\geq -\log_{10}\xi_0+4.2$, slightly enlarging the allowed parameter space compared to the classical case, where $\log_{10}p\geq -\log_{10}\xi_0+4.4$. Since we fixed $\mu$ to a constant value, we provide the potential values only for $\tau$ values for which the $\mu=const$ choice is justified.

In the following calculations, the slow-roll parameters $\epsilon_V$ and $\eta_V$ were derived using the $V_{1-\textrm{loop}}(\tau)$ potential (\ref{V1loop_full}). Then the couplings were evaluated at $\mu_1$, and the inflaton field was rescaled as $\tau\rightarrow\exp\big(\Gamma(\mu_1)\big)\tau$. Following the same procedure as in Section \ref{Section:Tree_Level}, we computed the scalar spectral index $n_s$ and the tensor-to-scalar ratio $r_{0.002}$, focusing on $N_e=50$. The results for $N_e=60$ are nearly identical and, therefore, omitted for brevity. 

\begin{figure}[!t]
     \centering
     \begin{subfigure}[c]{0.49\textwidth}
         \centering
         \includegraphics[width=\textwidth]{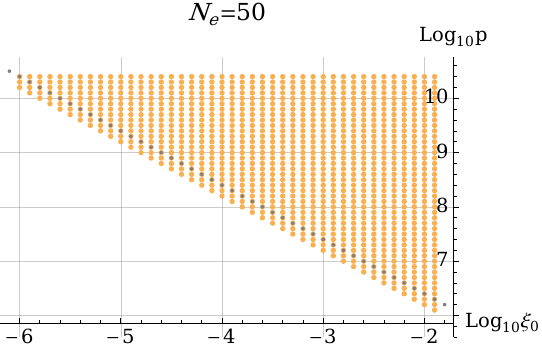}
    \end{subfigure} 
    \hfill
     \begin{subfigure}[c]{0.49\textwidth}
         \centering
         \includegraphics[width=\textwidth]{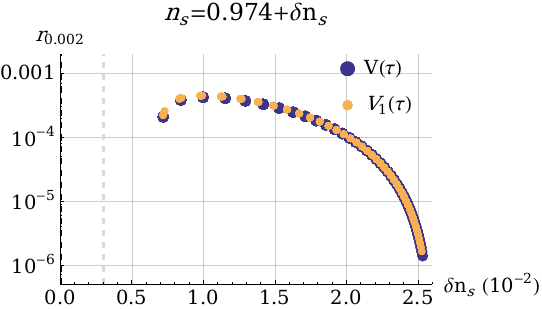}
     \end{subfigure}
    \caption{\textbf{Left:} Parameter space points used to derive the $r_{0.002}(n_s)$ function shown in the right panel. The dotted gray line marks the boundary of the tree-level constraint, given by $\log_{10}p\geq-\log_{10}\xi_0+4.4$. \textbf{Right:} The function $r_{0.002}(n_s)$, computed using the RG improved potential $V_{RGI}(\tau)$, for the endpoint of inflation defined as $\eta_V(\tau_{end})=-0.0095$ and $N_e=50$. The results are compared with the tree-level $r_{0.002}(n_s)$ function for the same parameters, as illustrated in Figure \ref{ns_r_indecies}.}
    \label{rns_plt_RGI}
\end{figure}

\begin{figure}[!t]
     \centering
         \centering
         \includegraphics[width=0.8\textwidth]{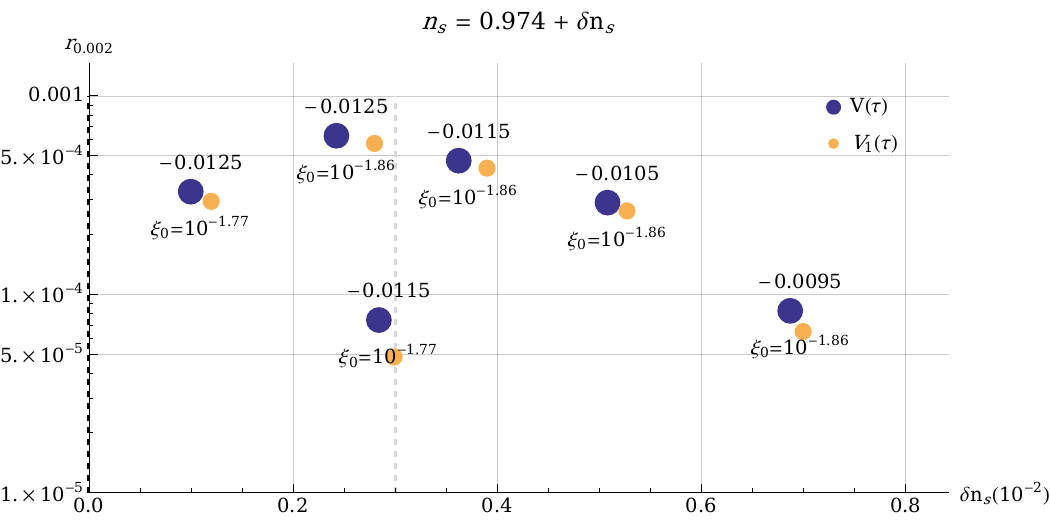}
    \caption{Tensor-to-scalar ratio $r_{0.002}(n_s)$ for $\xi_0=10^{-1.71}$ and $\xi_0=-1.86$ with $p=10^{6.5}$ at different inflationary endpoints, determined by $\eta_V(\tau_{end})$ and $N_e=50$. The numerical labels above the points indicate the corresponding $\eta_V(\tau_{end})$ values. The black dashed line corresponds to the experimental value, $n_s=0.974$, and the gray dashed line to $n_s=0.974+0.003$. The results are compared with the tree-level $r_{0.002}(n_s)$ function for the same parameters, as illustrated in Figure~\ref{rns_classical_diff_etaV}.}
    \label{rns_plt_RGI_diffeta}
\end{figure}

Figure \ref{rns_plt_RGI} illustrates the function $r(n_s)$ for both the tree-level and quantum-corrected potentials, assuming the end of inflation is defined by $\eta_{V}(\tau_{end})=-0.0095$. The figure also displays the parameter space points used in deriving these results. Again, the resulting values of $n_s$ do not satisfy the experimental bounds from Table \ref{tab:inflation_params}. The impact of $\xi_0$ and $p$ on $n_s$ and $r_{0.002}$ for $\eta_V(\tau_{end})=-0.0095$ follows the same pattern as in the tree-level analysis, as shown in Figure \ref{ns_r_contours}. Next, as in the tree-level case, we apply the analysis of the impact of the $\eta_V(\tau_{end})$ choice. Figure \ref{rns_plt_RGI_diffeta} illustrates this dependence for $\xi_0=10^{-1.86}$ and $\xi_0=10^{-1.77}$ with $p=10^{6.5}$, showing that the consistent results can be achieved. Finally, the quantum analogue of Figure \ref{allowed_class}, which would show the allowed parameter space for $n_s$, exhibits negligible differences and is therefore omitted.

\subsection{Gravitational Waves from Inflation}
\label{Subsection:GW_spectrum}

\begin{figure}[!t]
\vspace{-10pt}
\centering
\begin{subfigure}[c]{0.45\textwidth}
\vspace{40pt}
    \includegraphics[width = \textwidth]{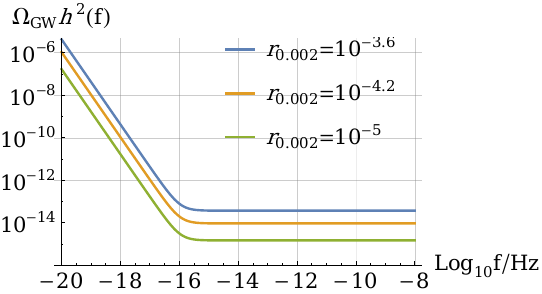}
    \caption{}
    \label{GW_results}
\end{subfigure}
\hfill
\begin{subfigure}[c]{0.51\textwidth}
    \includegraphics[width = \textwidth]{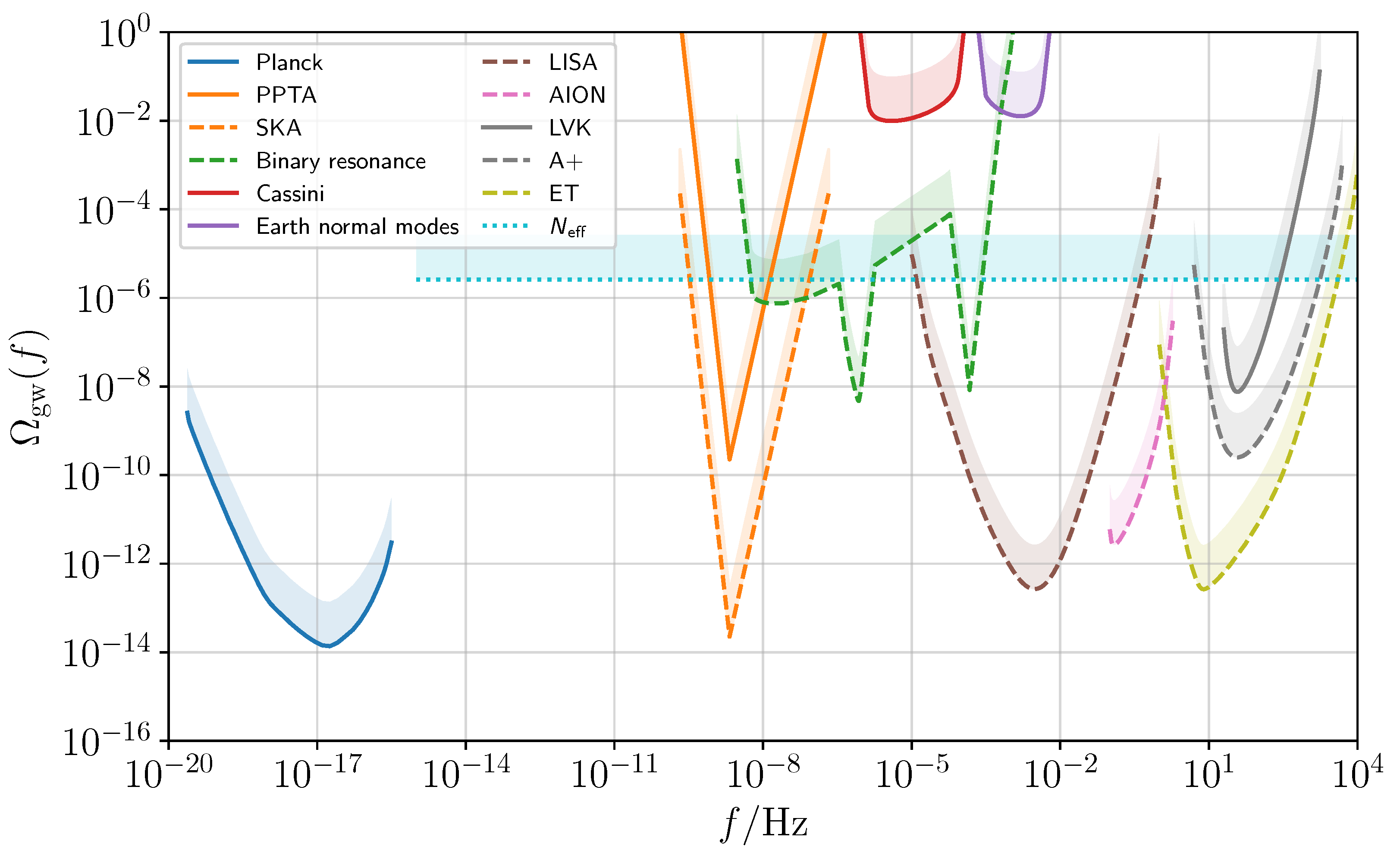}
    \caption{}
    \label{GW_experiments}
\end{subfigure}
\vspace{-10pt}
\caption{(a) Gravitational wave spectrum (\ref{GW_spectrum}) for three values of $r_{0.002}$ obtained from the analysis in Section \ref{Section:Quantum_Inflation}.(b) Detection sensitivity ranges of current gravitational wave experiments. Figure adapted from~\cite{Renzini:2022alw}.}
\vspace{-10pt}
\end{figure}

During inflation, quantum fluctuations in the gravitational field induce tensor metric perturbations, $h_{ij}$, representing small deviations from the background FLRW metric:
\begin{equation}
    ds^2 = a^2(\eta)\Big[-\textrm{d}\eta^2+\big(\delta_{ij}+h_{ij}\big)\textrm{d}x^i\textrm{d}x^j  \Big].
\end{equation}
As inflation progresses, these perturbations are amplified and stretched to super-Hubble scales, forming a nearly scale-invariant spectrum \cite{Caprini:2018mtu, Lasky:2015lej}. Once they re-enter the Hubble horizon after inflation, they evolve as classical gravitational waves (GWs), generating a stochastic background that could potentially be observed by experiments such as LISA, ET, CMB B-mode polarization searches, and future gravitational wave observatories. 

The present-day energy density spectrum of these inflationary GWs, $\Omega_{GW}(f)$, is given by \cite{Caprini:2018mtu, Lasky:2015lej}:
\begin{equation}
    \Omega_{GW}(f) = \frac{3}{128} r_{0.002} A_s \Omega_r \bigg(\frac{f}{f_*}\bigg)^{n_t}\Bigg[\frac{1}{2}\bigg(\frac{f_{eq}}{f}\bigg)^2+\frac{16}{9}\Bigg],
    \label{GW_spectrum}
\end{equation}
where $A_s$ is the scalar amplitude\footnote{The Planck 2018 reports:
\begin{equation}
\ln\big(10^{10}A_s\big) = 3.044\pm 0.014.
        \label{A_s_value}
\end{equation}}, $\Omega_r = 3.72\cdot 10^{-5}$ is the present-day radiation energy density, and $f_*$ is the pivot frequency:
\begin{equation}
    f_* = \frac{0.05\textrm{ Mpc}^{-1}}{2\pi}\approx 7.75\cdot 10^{-16} \textrm{ Hz}.
\end{equation}
The quantity $f_{eq}$ corresponds to the frequency associated with the matter-radiation equality epoch, approximately given by $f_{eq}\approx 2.02\cdot 10^{-16}$. The tensor spectral index, $n_t$, is related to the slow-roll parameter $\epsilon_V$ at first order via $n_t=-2\epsilon_V.$ Evaluating $\epsilon_V$ at $\tau_*$, $n_t$ can also be expressed in terms of $r_{0.002}$ as:
\begin{equation}
     n_t = -\frac{r_{0.002}}{8}.
\end{equation}
The cosmological parameter values used in this analysis are consistent with experimental data \cite{ACT:2025fju, ACT:2025tim} and references \cite{Caprini:2018mtu, Lasky:2015lej}. 

Figure \ref{GW_experiments} illustrates the sensitivity ranges of current GW detection experiments across different frequencies. The lowest frequencies, detectable by CMB B-mode polarization searches of the Planck satellite, correspond to the largest cosmological scales, tracking the earliest times and the highest energy scales of inflation.

Figure \ref{GW_results} presents the predicted GW spectrum for values of $r_{0.002}$ derived within the scale symmetric Higgs-dilaton model, as shown in Figure \ref{rns_plt_RGI_diffeta}. Comparing these results with the sensitivity ranges in Figure \ref{GW_experiments}, we conclude that the predicted GW signal lies within the detectable range, providing a potential test of the framework developed in this work.

%% file: Unitarity.tex
\section{Unitarity Bounds}
\label{Section:Unitarity}

In the Einstein frame, the gravity is in its canonical form. Therefore, the unitarity cutoff resulting from the interactions involving gravitons is of the order of the Planck mass. However, the SM gauge sector changes due to the conformal transformation (\ref{ConfTrans}) and changing to the polar coordinates (\ref{polar_coordinates}). This can induce unitarity violation above some energy scale, which we aim to find below.

\begin{figure}[!t]
    \centering
    \includegraphics[width=0.8\linewidth]{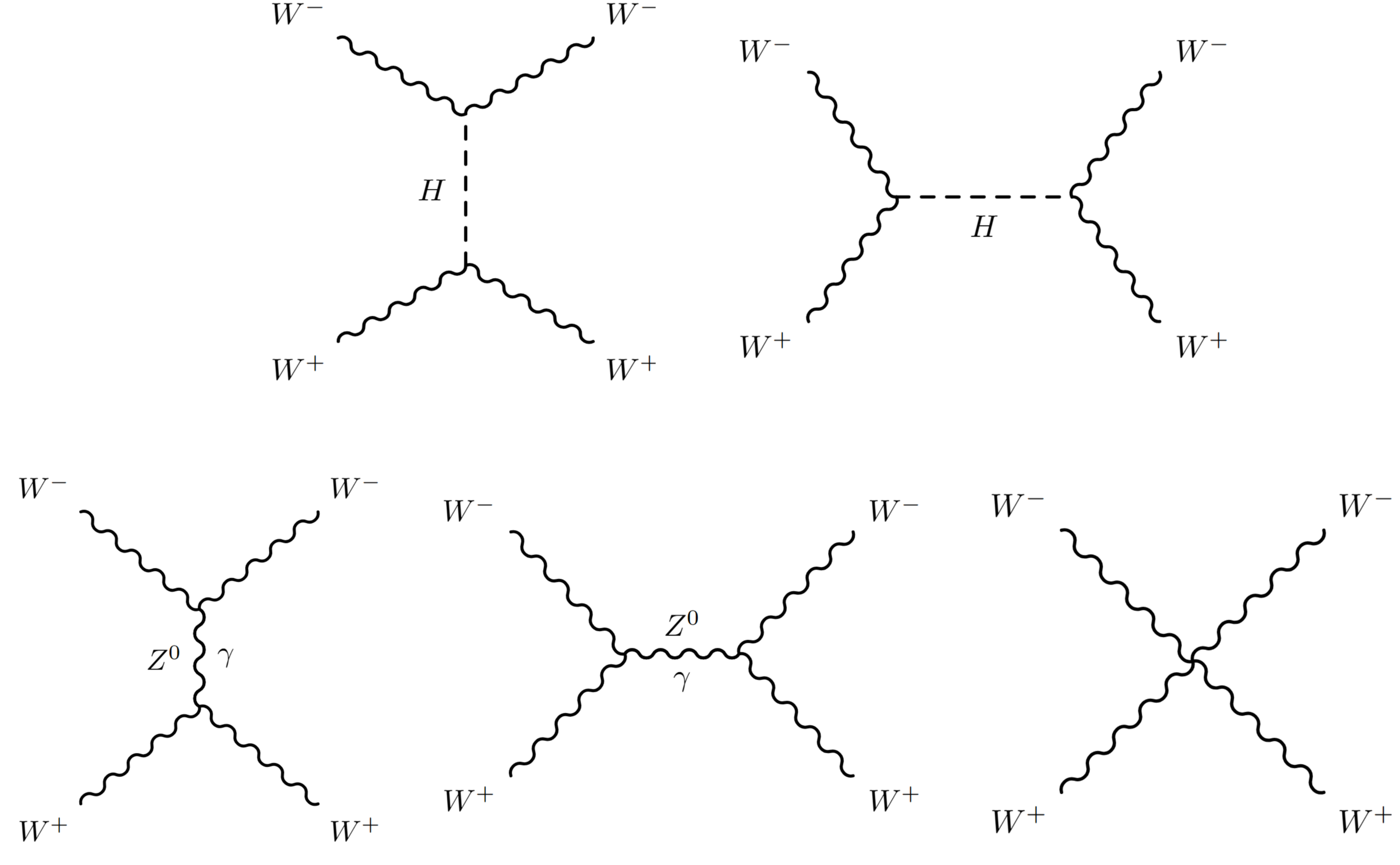}
    \caption{Diagrams contributing to $W_L^+W_L^-\rightarrow W_L^+W_L^-$ scattering amplitude.}
    \label{WW_H_diagrams}
\end{figure}

We analyze the behavior of scattering amplitudes for specific processes, particularly the longitudinal $W$ boson scattering $W_L^+W_L^-\rightarrow W_L^+W_L^-$. Figure \ref{WW_H_diagrams} illustrates the relevant Feynman diagrams contributing to this process. In the high-energy limit ($s, t\gg m_W^2$), the leading contributions from the $s-$ and $t-$channel exchange of the photon and $Z^0$, along with the 4-point gauge interaction give \cite{Cheung:2008zh,Szleper:2014xxa, Valencia:1990jp}:
\begin{equation}
    \mathcal{A}\big(W_L^+W_L^-\rightarrow W_L^+W_L^-\big)\Big|_{\textrm{gauge}} = -\frac{1}{4}g^2_2\frac{1}{m_W^2}u+\mathcal{O}(1),
    \label{WW_gauge_ampl}
\end{equation}
where $u$ is the Mandelstam variable, $u = 4m_W^2-s-t$. This amplitude grows with energy, signaling a potential violation of unitarity at high energies. However, the Higgs boson exchange in the $s$ and $t$ channels introduces a contribution of \cite{Cheung:2008zh, Szleper:2014xxa, Valencia:1990jp}:
\begin{equation}
    \mathcal{A}\big(W_L^+W_L^-\rightarrow W_L^+W_L^-\big)\Big|_{\textrm{Higgs}} = +\frac{1}{4}g^2_2\frac{1}{m_W^2}u+\mathcal{O}(1),
\end{equation}
which precisely cancels the problematic energy-growing term in the SM, ensuring a well-behaved amplitude in the high-energy limit. However, in the theories where the Higgs-$W$ interaction is modified, such as our model, this cancellation may not occur. 

We now examine the unitarity constraints in the Einstein frame Lagrangian of inflationary theory presented in this work. The charged $W^{\pm}$ bosons couple to the neutral Higgs component $\phi_1$ through the interaction term:
\begin{equation}
    \frac{\mathcal{L}_{Higgs-W}}{\sqrt{g}} \sim \frac{g_2^2}{4}g^{\mu\nu}W^+_{\mu}W^-_{\nu}\cdot\phi_1^2.
\end{equation}
After applying a conformal transformation and changing to polar coordinates, this becomes:
\begin{equation}
    \frac{\hat{\mathcal{L}}_{Higgs-W}}{\sqrt{\hat{g}}} \sim \frac{1}{\Omega^2}\frac{g_2^2}{4}\hat{g}^{\mu\nu}W^+_{\mu}W^-_{\nu}\cdot\frac{1}{1+\xi_1}K\cdot\frac{\tau^2}{1+\tau^2},
\end{equation}
where $\Omega^2$ is defined in equation (\ref{Omega_definition}). $K$ is an arbitrary scale, chosen to obtain proper vacuum expectation values for $\phi_i$ fields in equation (\ref{polar_coordinates}):
\begin{equation}
    K = \Bigg(-\frac{2\lambda_2}{\lambda_1}\big(1+\xi_0\big)+\big(1+\xi_1\big)\Bigg)\langle\phi_1^2\rangle, \qquad \langle\phi_1\rangle=250\textrm{ GeV}.
\end{equation}
Consequently, the Higgs$-W$ interaction is modified to:
\begin{equation}
    \frac{\hat{\mathcal{L}}_{Higgs-W}}{\sqrt{\hat{g}}}\sim \frac{g_2^2}{4}\hat{g}^{\mu\nu}W^+_{\mu}W^-_{\nu}\cdot f(\tau),
    \label{Higgs-W_interaction}
\end{equation}
where:
\begin{equation}
    f(\tau) = \frac{1}{\Omega^2}\frac{K}{1+\xi_1}\frac{\tau^2}{1+\tau^2} = 6M^2\frac{(1+\xi_0)\tau^2}{\xi_0(1+\xi_1)+\xi_1(1+\xi_0)\tau^2}.
\end{equation}
With the non-canonical kinetic term for the inflaton field:
\begin{equation}
    \frac{1}{2}F^2(\tau)\hat{g}^{\mu\nu}\partial_{\mu}\tau\partial_{\nu}\tau = \frac{1}{2}\hat{g}^{\mu\nu}\partial_{\mu}\sigma\partial_{\nu}\sigma, \qquad F(\tau) = \frac{\partial\sigma}{\partial\tau},
\end{equation}
we expand (\ref{Higgs-W_interaction}) around the $\tau_0$ background, with $\tau = \tau_0+\delta\tau$ and $\delta\tau\cdot F(\tau_0) = \delta\sigma$:
\begin{equation}
    \frac{\hat{\mathcal{L}}_{Higgs-W}}{\sqrt{\hat{g}}}\sim \frac{g_2^2}{4}\hat{g}^{\mu\nu}W^+_{\mu}W^-_{\nu}\cdot\Bigg(f(\tau_0)+\frac{f'(\tau_0)}{F(\tau_0)}\delta\sigma+\frac{f''(\tau_0)}{2F^2(\tau_0)}\delta\sigma^2+\ldots\Bigg).
\end{equation}
Consequently, the $WW\sigma$ vertex with canonical $\sigma$ field reads:
\begin{equation}
    g_{WW\sigma} = \frac{g_2^2}{4}\frac{f'(\tau_0)}{F(\tau_0)}.
\end{equation}
\begin{figure}[!t]
    \centering
    \includegraphics[width=0.5\linewidth]{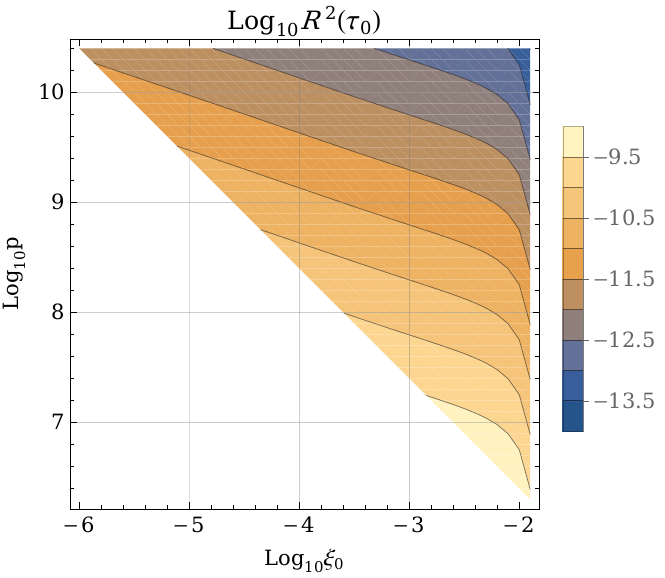}
    \caption{Plot of the $R^2(\tau_0)$ parameter, defined in (\ref{R2tau0}). The $\tau_0$ values for each plot point were determined as a $\tau^*$ for a corresponding parameter space point.}
    \label{R2tau0_plot}
\end{figure}
Defining the relative change of the interaction as:
\begin{equation}
    R(\tau_0) = \frac{g_{WW\sigma}}{g_2m_W(\tau_0)} = \frac{1}{2}\frac{f'(\tau_0)}{F(\tau_0)\sqrt{f(\tau_0)}},
    \label{R2tau0}
\end{equation}
the amplitude for the Higgs exchange now reads:
\begin{equation}
    \mathcal{A}\big(W_L^+W_L^-\rightarrow W_L^+W_L^-\big)\Big|_{\textrm{Higgs}} \simeq +\frac{1}{4}g^2_2\frac{R^2(\tau_0)}{m_W^2}u +\mathcal{O}(1).
\end{equation}
In the SM, $R=1$ and the Higgs exchange process cancels the amplitude in (\ref{WW_gauge_ampl})\footnote{The pure Yang-Mills sector is invariant under transformation to the Einstein frame.}. In the case of the theory presented in this paper, we examine the cancellation of amplitudes and the resulting unitary cutoff by the sum of both amplitudes:
\begin{equation}
    \mathcal{A}\big(W_L^+W_L^-\rightarrow W_L^+W_L^-\big) \simeq \frac{g_2^2}{4}\frac{u}{m_W^2(\tau_0)}\Big(-1+R^2(\tau_0)\Big).
\end{equation}
Figure \ref{R2tau0_plot} presents the $R^2(\tau_0)$ values for different $\xi_0$, $p$ and $\tau_0=\tau^*$ for a given parameter space with $\eta_V(\tau_{end})=-0.0095$ and $N_e=50$. Values are significantly smaller than 1. Therefore, the unitarity cutoff in the inflationary background is dominated by the gauge bosons' contribution:
\begin{equation}
    \Lambda_{UV}\Big|_{R\ll 1} \sim \sqrt{8\pi}\frac{m_W(\tau_0)}{g_2/2} = 8\sqrt{3\pi}M_P\tau_0\sqrt{\frac{1+\xi_0}{\xi_0(1+\xi_1) +\xi_1(1+\xi_0)\tau_0^2}}\xrightarrow[\quad\xi_0\ll 1, \tau_0 \lesssim 1\quad]{\xi_1\gg 1}\frac{M_P}{\sqrt{\xi_1}}.
\end{equation}
The cutoff always exceeds the values of inflationary potential $V(\tau)^{1/4}$. For a background value $\tau_0=\tau^*$, the cutoff exceeds the inflationary energy scale $V(\tau^*)^{1/4}$ by a factor of 11.9236, ensuring the validity of our results. The asymptotic behavior $M_P/\sqrt{\xi_1}$ matches the cutoff in the inflationary regime found in Higgs inflation models with a non-minimal coupling in the inflationary large field background \cite{Ito:2021ssc, Karananas:2022byw}.

\begin{figure}[!t]
     \centering
     \begin{subfigure}[c]{0.49\textwidth}
         \centering
         \includegraphics[width=\textwidth]{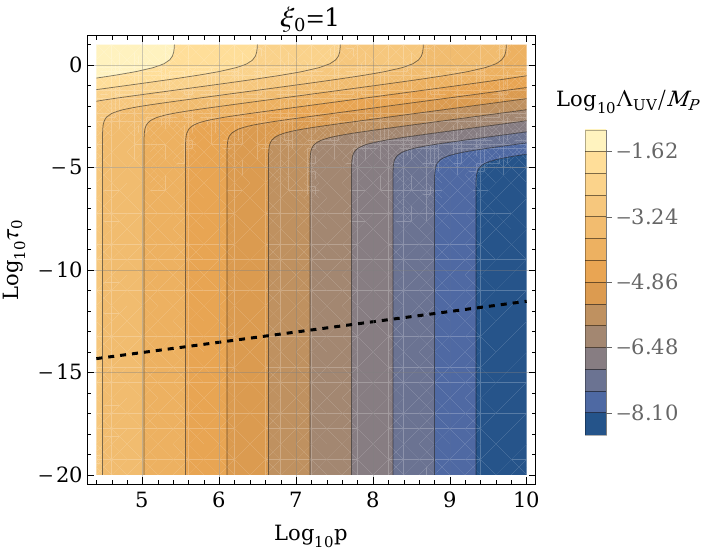}
    \end{subfigure} 
    \hfill
     \begin{subfigure}[c]{0.49\textwidth}
         \centering
         \includegraphics[width=\textwidth]{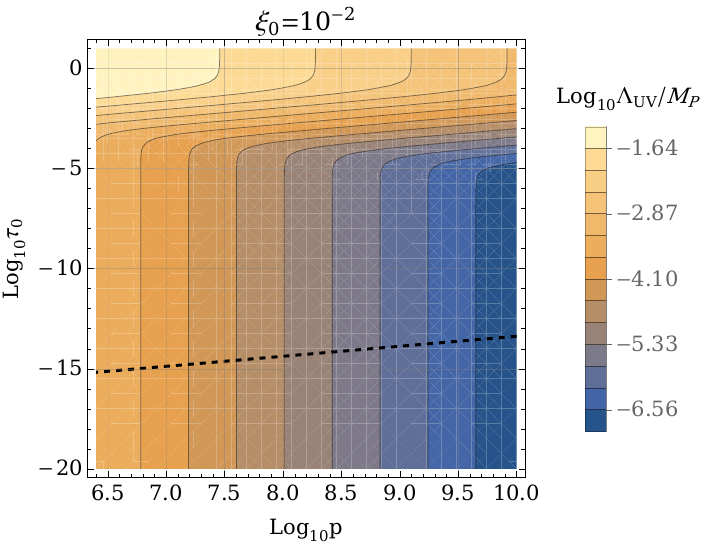}
     \end{subfigure}
    \caption{Cutoff scale $\Lambda_{UV}$ as a function of $\tau_0$ for different parameter-space points. The black dashed line indicates the vacuum expectation value $\langle\tau_0\rangle$.}
    \label{lambda_x0}
\end{figure}

The full $\tau_0$-dependent cutoff reads: 
\begin{equation}
    \Lambda_{UV} = \sqrt{8\pi}\frac{m_W(\tau_0)}{g_2/2}\frac{1}{\sqrt{-1+R^2(\tau_0)}},
\end{equation}
and is illustrated in Figure \ref{lambda_x0}. In the vacuum limit ($\tau_0 \ll 1$), the cutoff is significantly lower than in the inflationary background, in particular:
\begin{equation}
    \Lambda_{UV}\xrightarrow[\quad\xi_0\ll 1, \tau_0 \ll 1\quad]{\xi_1\gg 1}\frac{M_P}{\xi_1}.
    \label{asympt_cutoff_vacuum}
\end{equation}
The asymptotic behavior (\ref{asympt_cutoff_vacuum}) agrees with the Higgs inflation models from the literature~\cite{Ito:2021ssc, Karananas:2022byw}.

%% file: Conclusions.tex
\section{Conclusions}

In this work, we explored the inflationary dynamics of a scale-symmetric extension of the SM Higgs scalar sector, incorporating a dilaton field within the framework of Weyl geometry, where both scalars are non-minimally coupled to gravity. At both classical and quantum levels, the model supports slow-roll inflation, with the inflaton field $\tau$ exhibiting a non-canonical kinetic term and a nearly flat potential in the large-field regime. The model's predictions for inflationary observables, including the tensor-to-scalar ratio $r_{0.002}$ and the spectral index $n_s$, are generally consistent with the experimental constraints, although minor adjustments to the inflationary endpoint may be required to fully align with observational data. Upon inclusion of quantum corrections, the model incorporates propagator suppression factors, which modify the renormalization group equations and affect the running of couplings. Notably, these effects help stabilize the Higgs quartic coupling $\lambda_2$, preventing it from becoming negative at high energies.

Despite these advancements, several open questions remain. Most importantly, the inflationary analysis assumes that the Weyl vector mass $m_{\omega}\sim q\cdot M_P$ is at least of the Planck mass order, allowing us to neglect some of the $\omega_{\mu}$ effects on the theory. However, for smaller values of the gauge coupling $q$, this approximation breaks down, and the corresponding effects require careful study. Moreover, the present analysis has focused primarily on inflationary observables, leaving open questions regarding reheating, the generation of the baryon asymmetry, and possible dark matter candidates within the same scale-symmetric framework. Finally, while the predicted gravitational-wave spectrum suggests potential detectability, a more systematic study of post-inflationary dynamics is needed to establish whether the signal can serve as a distinctive probe of this model.

In summary, this work contributes to the ongoing exploration of scale symmetry and Weyl geometry in cosmology, offering insights into inflationary dynamics. However, further research is needed to fully realize the potential of this framework and address the remaining challenges in aligning theoretical predictions with observational constraints.

%% file: Appendices.tex
\appendixtitleon
\appendixtitletocon

\begin{appendices}

\section{Weyl geometry}
\label{Appendix:Weyl_geometry}

In this appendix, we outline the basic properties of Weyl's conformal geometry, which has been broadly studied in literature \cite{1, 2, 3,  TANN, ODA1, SCHOLZ, ODA3, ODA4, ODA5, Ghilencea:2019rqj, Ghilencea:2020rxc, Ghilencea:2021lpa, Ghilencea:2021jjl, Ghilencea:2022lcl}. We construct an action for the Higgs neutral component, $\phi_1$, and the dilaton, $\phi_0$, ensuring invariance under Weyl conformal transformations and, consequently, scale symmetry. Weyl quantities are denoted with a tilde, while their Riemannian counterparts remain plain. We adopt the metric convention $(+,-,-,-)$, with $g = |\det g_{\mu\nu} |>0$. 

In Weyl geometry, in contrast to the Riemannian geometry, where $\nabla _{\mu}g_{\alpha\beta} = 0$, the Weyl covariant derivative of the metric is non-zero, $\Tilde{\nabla}_{\mu}g_{\alpha\beta}=-\omega_{\mu}g_{\alpha\beta}$. The $\omega_{\mu}$ is the Weyl vector field, which, as a consequence, modifies geometric structures. The Weyl connection $\Tilde{\Gamma}^{\rho}_{\mu\nu}$ is torsion-free, i.e., $\Tilde{\Gamma}^{\rho}_{\mu\nu}=\Tilde{\Gamma}^{\rho}_{\nu\mu}$, and is given by:
\begin{equation}
    \Tilde{\Gamma}^{\rho}_{\mu\nu} = \Gamma^{\rho}_{\mu\nu}+\frac{q}{2}\Big[\delta^{\rho}_{\mu}\omega_{\nu}+\delta^{\rho}_{\nu}\omega_{\mu}-g_{\mu\nu}\omega^{\rho}\Big],
    \label{TildeGamma}
\end{equation}
where $q$ is the Weyl gauge coupling. As a result, the Weyl curvature scalar takes the form:
\begin{equation}
    \Tilde{R} = R -3q\nabla_{\mu}\omega^{\mu}-\frac{3}{2}q^2\omega_{\mu}\omega^{\mu},
    \label{Tilde_R_def}
\end{equation}
where $\nabla_{\mu}$ is the Riemannian covariant derivative with the Levi-Civita connection:
\begin{equation}
    \nabla_{\mu}\omega_{\alpha} = \partial_{\mu}\omega_{\alpha} - \Gamma^{\sigma}_{\mu\alpha}\omega_{\sigma}.
\end{equation}
The field strength tensor $\Tilde{F}_{\mu\nu}$ for the Weyl vector field remains identical to its Riemannian counterpart:
\begin{equation}
    \Tilde{F}_{\mu\nu} = \Tilde{\nabla}_{\mu}\omega_{\nu}-\Tilde{\nabla}_{\nu}\omega_{\mu} = \partial_{\mu}\omega_{\nu}-\partial_{\nu}\omega_{\mu} = F_{\mu\nu}.
    \label{Fmunu_Weyl}
\end{equation}

The Weyl gauge conformal transformations for the metric $g_{\mu\nu}$, scalar field $\phi$ and Weyl vector field $\omega_{\mu}$ are given by:
\begin{equation}
    \begin{split}
        g_{\mu\nu}\quad & \longrightarrow \quad g'_{\mu\nu} = \Omega^2g_{\mu\nu}, \\
        \phi \quad & \longrightarrow \quad \phi' = \frac{\phi}{\Omega}, \\
        \omega_{\mu} \quad & \longrightarrow \quad \omega'_{\mu} = \omega_{\mu} - \frac{1}{q}\partial_{\mu}\ln\Omega^2,
    \end{split}
    \label{ConfTrans}
\end{equation}
where $\Omega$ is a dimensionless real parameter that may, in general, be field-dependent. Consequently, we obtain $g^{\prime\mu\nu} = \Omega^{-2}g^{\mu\nu}$, $\sqrt{g'} =\Omega^4\sqrt{g}$ and $\Tilde{R}' = \Omega^{-2}\Tilde{R}$. Unlike the Riemannian connection $\Gamma^{\sigma}_{\mu\nu}$, the Weyl connection $\Tilde{\Gamma}^{\sigma}_{\mu\nu}$ remains invariant under (\ref{ConfTrans}). Additionally, the Weyl field strength tensor remains unchanged, $\Tilde{F}_{\mu\nu}' = \Tilde{F}_{\mu\nu}$.

The action for two scalar fields, the dilaton $\phi_0$ and the Higgs neutral component $\phi_1$, embedded in Weyl geometry and invariant under (\ref{ConfTrans}), is given by:
\begin{equation}
    \frac{\mathcal{L}}{\sqrt{g}} = -\frac{1}{12}\Big(\xi_0\phi_0^2+\xi_1\phi_1^2\Big)\Tilde{R}-\frac{1}{4}\Tilde{F}_{\mu\nu}\Tilde{F}^{\mu\nu}+\frac{1}{2}\Tilde{D}_{\mu}\phi_0\Tilde{D}^{\mu}\phi_0+ \frac{1}{2}\Tilde{D}_{\mu}\phi_1\Tilde{D}^{\mu}\phi_1-V(\phi_0,\phi_1),
\end{equation}
where the Weyl-covariant derivative:
\begin{equation}
    \Tilde{D}_{\mu}\phi_i = \Big(\partial_{\mu}-\frac{q}{2}\omega_{\mu}\Big)\phi_i,
    \label{Tilde_Dmu}
\end{equation}
transforms under (\ref{ConfTrans}) as the scalar field:
\begin{equation}
    \Tilde{D}_{\mu}'\phi' = \Omega^{-1}\Tilde{D}_{\mu}\phi.
\end{equation}
To ensure invariance under conformal transformations, the potential $V(\phi_0,\phi_1)$ must be a homogeneous function of degree four, leading to:
\begin{equation}
    V(\phi_0,\phi_1) = \lambda_0\phi_0^4+\lambda_1\phi_0^2\phi_1^2+\lambda_2\phi_1^4.
\end{equation}

\section{Parameter Space Constraints}
\label{Appendix:Parameter_space}

Here, we derive the parameter space constraints resulting from the scale-symmetric tree-level model of the Higgs-dilaton theory. As the experimental bounds refer to the energies around the electroweak scale, the Weyl vector field $\omega_{\mu}$ is decoupled and the model is embedded in Riemannian geometry with Ricci curvature scalar $R$. The Lagrangian for the Higgs neutral component $\phi_1$ and the dilaton $\phi_0$ reads:
\begin{equation}
    \frac{\mathcal{L}}{\sqrt{g}} = -\frac{1}{12}\Big(\xi_0\phi_0^2+\xi_1\phi_1^2\Big)R+\frac{1}{2}\partial_{\mu}\phi_0\partial^{\mu}\phi_0+\frac{1}{2}\partial_{\mu}\phi_1\partial^{\mu}\phi_1-V(\phi_0,\phi_1),
    \label{Llowlimit}
\end{equation}
with the scale symmetric potential:
\begin{equation}
    V(\phi_0,\phi_1) = \lambda_0\phi_0^4+\lambda_1\phi_0^2\phi_1^2+\lambda_2\phi_1^4.
    \label{Vtreelevel}
\end{equation}
The coupling constants satisfy certain hierarchy:
\begin{equation}
    \lambda_2\gg |\lambda_1| \gg \lambda_0, \qquad \lambda_0>0, \lambda_1<0, \lambda_2>0,
    \label{Lambdas_hierarchy}
\end{equation}
so that the new dilaton sector is weakly coupled. Stationary solutions to equations of motion along with a zero cosmological constant condition at the ground state:
\begin{equation}
    V(\langle\phi_0\rangle,\langle\phi_1\rangle)=0,
\end{equation}
result in the flat direction in the field space:
\begin{equation}
    \langle\phi_1^2\rangle = -\frac{\lambda_1}{2\lambda_2}\langle\phi_0^2\rangle, \qquad \lambda_0=\frac{\lambda_1^2}{4\lambda_2},
    \label{Higgs_vev}
\end{equation}
which agrees with the scale-invariance of the theory. The mass matrix:
\begin{equation}
M^2 = 
\left( \begin{array}{cc}
\lambda _1 \left(2 \phi_1^2+\frac{3 \lambda _1 }{\lambda _2}\phi_0^2\right) & 4\lambda_1\phi_1\phi_0 \\
4\lambda_1\phi_1\phi_0 & 2 \left(6  \lambda _2\phi_1^2+\lambda _1 \phi_0^2\right)
\end{array}
\right),
\label{M2}
\end{equation}
has two eigenvalues at the ground state:
\begin{equation}
m_G^2 = 0, \qquad m_H^2 = -4\lambda_1\Big(1-\frac{\lambda_1}{2\lambda_2}\Big)\langle\phi_0^2\rangle.
\label{Higgs_mass}
\end{equation}

\begin{figure}[!t]
     \centering
     \begin{subfigure}[b]{0.48\textwidth}
     \centering
         \includegraphics[width=\textwidth]{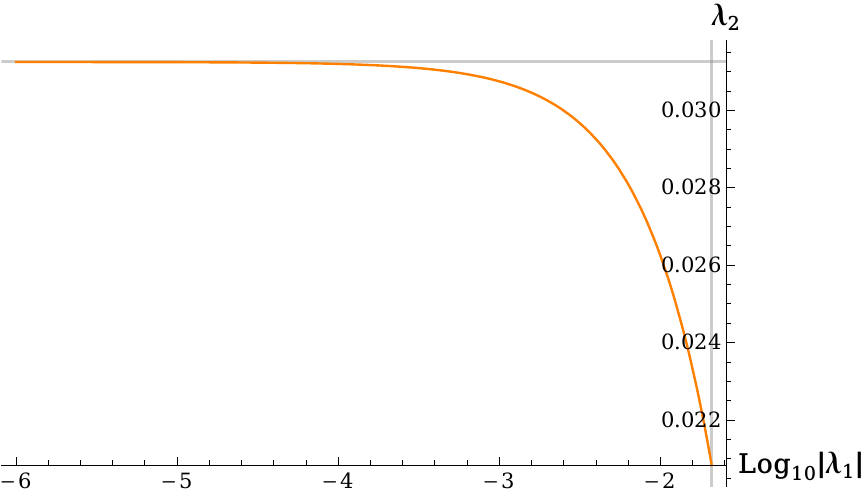}
     \end{subfigure}
     \hfill
     \begin{subfigure}[b]{0.48\textwidth}
         \centering
         \includegraphics[width=\textwidth]{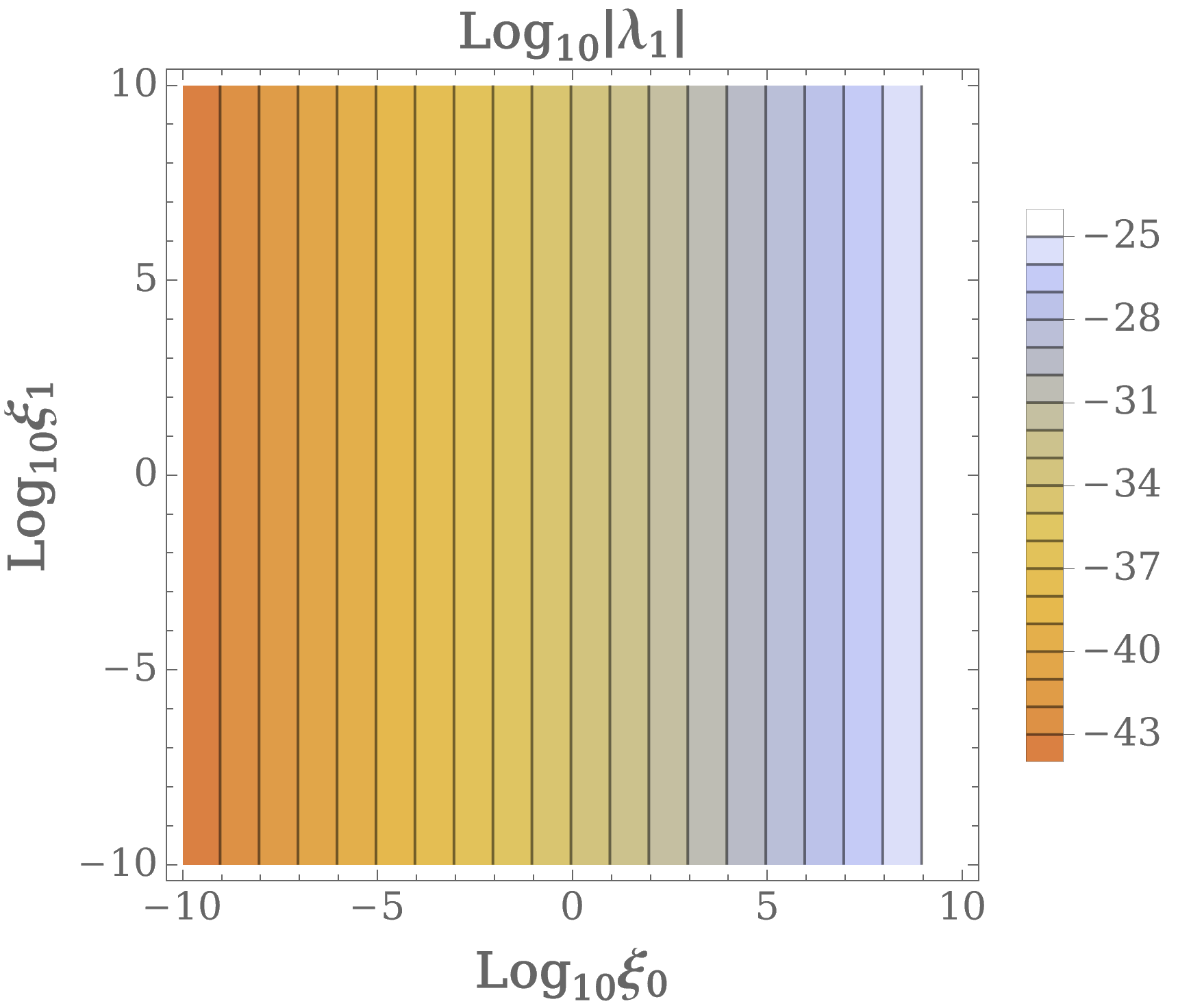}
    \end{subfigure} 
    \caption{Plots illustrating the allowed parameter space. Left: Relation $\lambda_2 = \frac{1}{32}\big(1+16\lambda_1\big)$ for possible $\lambda_1$ values. For sufficiently small $|\lambda_1|$, $\lambda_2$ tends to a constant value of 0.03125, marked by the gray grid line on the $y$ axis. The gray grid line on $x$ axis correspond to the boundary value $\lambda_1 = -\frac{1}{48}$. Right: Possible values of $\lambda_1$ as a function of non-minimal couplings $\xi_i$, according to relation (\ref{couplings_relations}). For $\xi_0$ values considered in this work, the resulting $|\lambda_1|$ indicates that the new dilaton sector is super-weakly coupled to the Higgs scalar.}
    \label{parameter_space}
\end{figure}

Spontaneous scale symmetry breaking (SSSB) occurs when the dilaton acquires its vev, and the flat direction no longer exists. After both fields acquire their vevs, the Ricci scalar term becomes the Einstein-Hilbert term:
\begin{equation}
    -\frac{1}{12}\Big(\xi_0\phi_0^2+\xi_1\phi_1^2\Big)R \quad \xrightarrow{\textrm{ SSSB }}\quad -\frac{1}{2}M_P^2 R,
\end{equation}
Consequently, the Planck mass scale is determined as:
\begin{equation}
    M_P^2 = \frac{1}{6}\Big(\xi_0\langle\phi_0^2\rangle+\xi_1\langle\phi_1^2\rangle\Big) = \frac{1}{6}\Big(\xi_0-\frac{\lambda_1}{2\lambda_2}\xi_1\Big) \langle\phi_0^2\rangle.
    \label{Planck_mass}
\end{equation}
The dilaton's vev $\langle\phi_0\rangle$ generates the fundamental scale parameters such as the Higgs vev $\langle\phi_1\rangle$, the Higgs mass $m_H$, and the Planck mass $M_{P}$. In scale-invariant theories, the value of $\langle\phi_0\rangle$ remains arbitrary, as only the relative ratios of mass scales can be established. The hierarchy (\ref{Lambdas_hierarchy}) leads to the relation $\langle\phi_0\rangle\gg\langle\phi_1\rangle$, which plays an important role in generating the $M_P$ and allows the dilaton vev to be of the order of the Planck mass. 

The possible parameter space of the theory can be obtained using the relations (\ref{Higgs_vev}), (\ref{Higgs_mass}), and (\ref{Planck_mass}), along with the hierarchy (\ref{Lambdas_hierarchy}). In the forthcoming calculations, we used the approximate experimental values \cite{pdg_astrophysical_constants_2021, pdg_higgs_boson_2023}:
\begin{equation}
    m_H = 125\textrm{ GeV},\qquad \langle\phi_1\rangle = 250\textrm{ GeV}, \qquad M_{P} = 2.44\cdot 10^{18}\textrm{ GeV}.
\end{equation}
Setting $\xi_1=p\cdot\xi_0$, the parameter space constraints are as follows:
\begin{equation}
    \lambda_2 = \frac{1}{32}\big(1+16\lambda_1\big), \qquad \frac{1}{\lambda_1} = -16\Bigg(\frac{6M_P^2}{\xi_0\langle\phi_1^2\rangle}-p+1\Bigg),
    \label{couplings_relations}
\end{equation}
\begin{equation}
    0>\lambda_1\gg-\frac{1}{48}\approx -0.021, \qquad \xi_0\big(2+p\big)\ll\frac{6M^2_P}{\langle\phi_1^2\rangle} \approx 5.72\cdot 10^{32}.
\end{equation}
Figure \ref{parameter_space} illustrates the couplings relations (\ref{couplings_relations}).

\section{Beta Functions}
\label{Appendix:Beta_functions}

\begin{figure}[!t]
     \centering
     \begin{subfigure}[c]{0.5\textwidth}
         \centering
         \includegraphics[width=\textwidth]{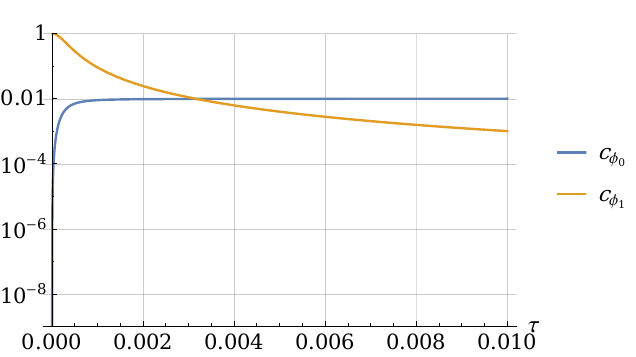}
     \end{subfigure}
     \hfill
     \begin{subfigure}[c]{0.45\textwidth}
         \centering
         \includegraphics[width=\textwidth]{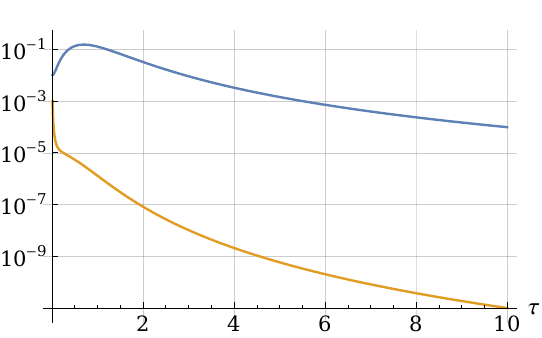}
    \end{subfigure} 
    \vspace{-5pt}
    \caption{Plots of the suppression factors $c_{\phi_i}(\tau)$ (\ref{c_factors_definition}), for $\xi_0=10^{-2}$ and $\xi_1=10^5$. Note the different $\tau$ scales on each plot. The low $\tau$ field value limits are $\lim_{\tau\rightarrow 0}c_{\phi_0} = 0$ and $\lim_{\tau\rightarrow 0}c_{\phi_1}=1$.}
    \label{c_factors figure}
\end{figure}

The canonical commutation relations $\big[\phi(x),\dot{\phi}(y) \big]$ are modified in models with non-minimal couplings to gravity for scalar fields, such as $\xi_i\phi_i^2 R$ terms. This modification has been previously discussed in literature \cite{Lerner:2009xg, Lee:2013nv, Salopek:1988qh, Lerner:2011ge}, which argumentation we follow. In the Einstein frame Lagrangian (\ref{Lagr_infl_eoms}), the scalar's kinetic term is non-canonical, and in terms of the inflaton field $\tau$, reads:
\begin{equation}
    \frac{\mathcal{L}_E}{\sqrt{g_E}}\Bigg|_{k.t.} = \frac{1}{2}F^2(\tau)g^{0\nu}_E\partial_{\nu}\tau\partial_0\tau = \frac{1}{2}g_E^{0\nu}\partial_{\nu}\sigma\partial_{0}\sigma,
\end{equation}
where the subscript $E$ denotes the quantities in the Einstein frame, $\sigma$ is the canonical field (\ref{canonical_field}) and $F^2(\tau) = \big(\partial\sigma/\partial\tau\big)^2$. In this frame, the gravity sector is canonical, and $\phi_i$ commutes with its canonical momentum $\pi_i$, i.e., $\big[ \phi_i(x),\pi_j(y)\big] = i\delta_{ij}\delta^{(3)}\big(\vec{x}-\vec{y}\big)$. The $\pi_i$, accounting for the frame change effects, is calculated as follows:
\begin{equation}
    \begin{split}
    \pi_i = & \frac{\partial\mathcal{L}_E\big|_{k.t.}}{\partial\dot{\phi_i}} = \sqrt{g_E}\frac{1}{2}g^{0\nu}_E\cdot \frac{\partial}{\partial\dot{\phi_i}}\Big(\partial_{\nu}\sigma\partial_{0}\sigma\Big) = \frac{1}{2}\Omega^2\sqrt{g}g^{0\nu} \cdot \frac{\partial}{\partial\dot{\phi_i}}\Big(\partial_{\nu}\sigma\partial_{0}\sigma\Big) = \\
    = & \frac{1}{2}\Omega^2\sqrt{g}\cdot \frac{\partial}{\partial\dot{\phi_i}}\dot{\sigma}^2 = \frac{1}{2}\Omega^2\sqrt{g}\cdot 2\dot{\sigma}\frac{\partial\sigma}{\partial\phi_i} = \Omega^2\sqrt{g}\Bigg(\frac{\partial\sigma}{\partial\phi_i}\Bigg)^2\dot{\phi}_i = \Omega^2\sqrt{g} F^2(\tau)\Bigg(\frac{\partial\tau}{\partial\phi_i}\Bigg)^2\dot{\phi}_i,
    \end{split}
\end{equation}
where Einstein frame metric functions have been transformed to the Jordan-frame ones, i.e., $\sqrt{g_E} = \Omega^4\sqrt{g}$ and $g_E^{\mu\nu}=\Omega^{-2}g^{\mu\nu}$, with $\Omega^2$ defined in (\ref{Omega_definition}). Then:
\begin{equation}
    \big[ \phi_i(x),\pi_i(y)\big] = \Omega^2\sqrt{g} F^2(\tau)\Bigg(\frac{\partial\tau}{\partial\phi_i}\Bigg)^2\big[\phi_i(x),\dot{\phi}_i(y)\big] = i\delta_{ij}\delta^{(3)}\big(\vec{x}-\vec{y}\big),
\end{equation}
from which follows:
\begin{equation}
    \big[\phi_i(x),\dot{\phi}_i(y)\big]  = \frac{i}{a^3}\cdot c_{\phi_i} \cdot \delta^{(3)}\big(\vec{x}-\vec{y}\big),
\end{equation}
where the suppression factors $c_{\phi_i}$ are defined as:
\begin{equation}
    c_{\phi_i} = \frac{1}{\Omega^2\Big(\frac{\partial\sigma}{\partial\phi_i}\Big)^2} = \frac{1}{\Omega^2 F^2(\tau)}\Bigg(\frac{\partial\phi_i}{\partial\tau}\Bigg)^2.
    \label{c_factors_definition_appendix}
\end{equation}
Figure \ref{c_factors figure} illustrates the suppression factors $c_{\phi_i}(\tau)$ for an example $\xi_i$ values. They suppress the $\phi_i$ fields propagators, in particular, in the large $\tau$ regime, when the inflation occurs. This affects the $\beta$ functions not only for the $\lambda_i$, $\xi_i$ coupling from the potential (\ref{Vtau_infl}), but also SM Couplings $g_1$, $g_2$ and $y_t$, present in $\lambda_i$ RG equations. 

Below, we provide the one-loop $\beta$ functions for couplings $\alpha_i = \{\lambda_0,\lambda_1,\lambda_2, \xi_0, \xi_1, g_1,g_2,g_3,y_t\}$, which can be found in \cite{Lerner:2009xg, Lee:2013nv, Lerner:2011ge} (note, that the notation used in this work may differ from that in the references). We set the notation $B_i = \big(4\pi\big)^2\cdot\beta_{i} = \big(4\pi\big)^2\cdot \frac{\textrm{d}}{\textrm{d}\ln\mu}\alpha_i$ and $c_i = c_{\phi_i}$. Setting $c_{\phi_i}= 1$ restores the usual $\beta$ functions, that can be found in \cite{Czerwinska:2015xwa, Lebedev:2021xey}.

\begin{equation}
    \begin{split}
        B_{\xi_0} & = 24c_0\lambda_0\big(\xi_0+1\big)+\big(3+c_1\big)\lambda_1\big(\xi_1+1\big),\\
        B_{\xi_1} & = 4c_0\lambda_1\big(\xi_0+1\big)+\bigg(12\big(1+c_1\big)\lambda_2-\frac{3}{2}\big(3g_1^2+g_2^2-4y_t^2\big)\bigg)\big(\xi_1+1\big),\\
        B_{\lambda_0} &= 72c_0^2\lambda_0^2+2\big(3+c_1^2\big)\lambda_1^2,\\
        B_{\lambda_1} &= 8\lambda_1\bigg(3c_0^2\lambda_0+2c_0c_1\lambda_1+3\big(1+c_1^2\big)\lambda_2\bigg)-3\lambda_1\big(3g_1^2+g_2^2-4y_t^2\big),\\
        B_{\lambda_2} & = 2c_0^2\lambda_1^2 + 24\big(1+3c_1^2\big)\lambda_2^2-6\lambda_2\big(3g_1^2+g_2^2-4y_t^2\big)+\frac{3}{32}\big((g_1^2+g_2^2)^2+2g_2^4\big)-\frac{3}{2}y_t^2,\\
        B_{g_1} & = \frac{40+c_1}{6}g_1^3, \\
        B_{g_2} & = -\frac{20-c_1}{6}g_2^3, \\
        B_{g_3} & = -7g_3^3, \\
        B_{y_t} & = y_t\Bigg(\frac{9}{2}c_1y_t^2-\frac{17}{12}g_1^2-\frac{9}{4}g_2^2-8g_2^2\Bigg).
\end{split}
\label{all_beta_functions}
\end{equation}
The initial values $\alpha_i\big(\mu_0\big)$, with $\mu_0=\langle\phi_1\rangle=250\textrm{ GeV}$, are given as:
\begin{equation}
\begin{split}
    &I_{\lambda_0} = \frac{I_{\lambda_1}^2}{4I_{\lambda_2}}, \quad I_{\lambda_1} = -\frac{1}{16}\Bigg(\frac{6M_P^2}{I_{\xi_0}\langle\phi_1^2\rangle}-p+1 \Bigg)^{-1}, \quad I_{\lambda_2} = \frac{1}{32}\big(1+16I_{\lambda_1}\big), \quad I_{\xi_1} = p\cdot I_{\xi_0}, \\
    &\qquad\qquad\qquad  I_{g_1} = 0.35940, \quad I_{g_2}=0.64754, \quad I_{y_t} = 0.95113,
    \label{init_couplings}
\end{split}
\end{equation}
where the individual relations come from the parameter space analysis in Appendix \ref{Appendix:Parameter_space}.

\begin{figure}[!t]
      \centering
     \begin{subfigure}[c]{\textwidth}
         \centering
         \includegraphics[width=0.3\textwidth]{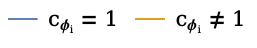}
     \end{subfigure}
     \\
     \begin{subfigure}[c]{0.5\textwidth}
         \centering
         \includegraphics[width=\textwidth]{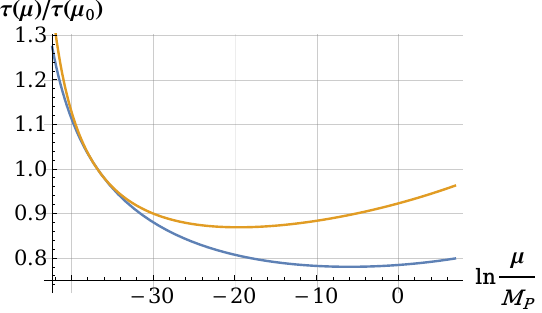}
    \end{subfigure} 
    \vspace{-5pt}
    \caption{The running of the inflaton field $\tau$ with the energy scale $\mu$, according to equation (\ref{tau_run_eq}).}
    \label{tau_run_plt}
    \vspace{-5pt}
\end{figure}

The running of the inflaton field is determined by the anomalous dimensions of the Higgs field $\gamma_H = \frac{d\ln \phi_1}{d\ln\mu}$ and the dilaton field $\gamma_G = \frac{d\ln \phi_0}{d\ln\mu}$, which are equal to \cite{LOG, Einhorn:2007rv, Buttazzo:2013uya}:
\begin{equation}
    \gamma_H = \frac{2}{\big(4\pi\big)^2}\Bigg(\frac{9}{4}g_1^2+\frac{3}{4}g_2^2-3y_t^2\Bigg), \qquad \gamma_G=0.
\end{equation}
Consequently, we obtain:
\begin{equation}
    \frac{d\ln \tau}{d\ln \mu} = \frac{d}{d\ln\mu}\sqrt{\frac{1+\xi_1}{1+\xi_0}}\frac{\phi_1}{\phi_0} = \frac{d\ln \phi_1}{d\ln\mu} = \gamma_H.
    \label{tau_run_eq}
\end{equation}
To RG improve the potential (\ref{V1loop_full}), $V_{1-\textrm{loop}}\Big|_{RGI} = V_{RGI}(\tau)$, we make the replacement $\tau\rightarrow\exp\big(\Gamma(\mu)\big)\tau$, where:
\begin{equation}
    \Gamma(\mu) = \int_{\langle\phi_1\rangle}^{\mu} \gamma_H(\mu')d\ln\mu'.
    \label{Gamma_mu}
\end{equation}

\begin{figure}[!t]
     \centering
     \begin{subfigure}[c]{\textwidth}
         \centering
         \includegraphics[width=0.3\textwidth]{cflegend.pdf}
     \end{subfigure}
     \\
     \begin{subfigure}[c]{0.48\textwidth}
         \centering
         \includegraphics[width=\textwidth]{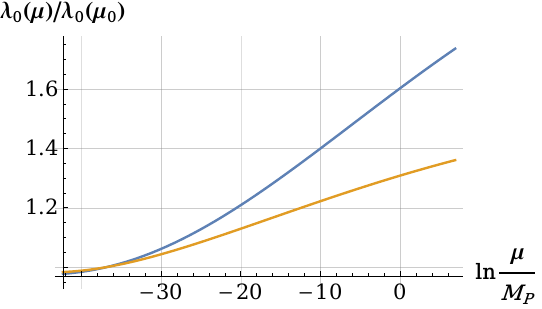}
     \end{subfigure}
     \hfill
     \begin{subfigure}[c]{0.48\textwidth}
         \centering
         \includegraphics[width=\textwidth]{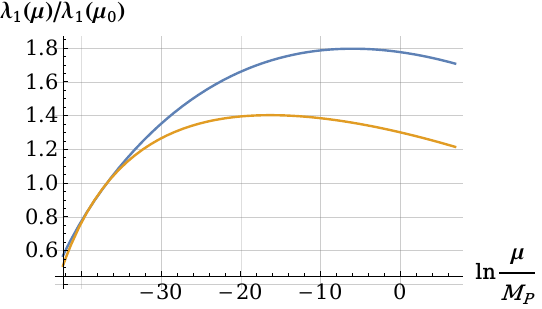}
    \end{subfigure} 
    \vspace{5pt}
    \\
     \begin{subfigure}[c]{0.48\textwidth}
         \centering
         \includegraphics[width=\textwidth]{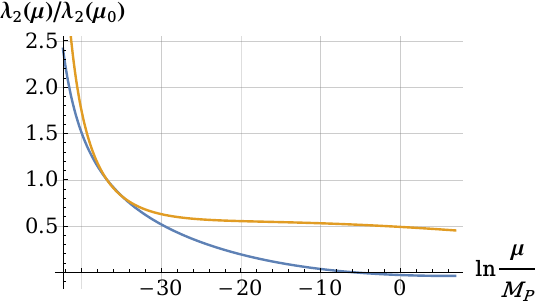}
     \end{subfigure}
     \hfill
     \begin{subfigure}[c]{0.48\textwidth}
         \centering
         \includegraphics[width=\textwidth]{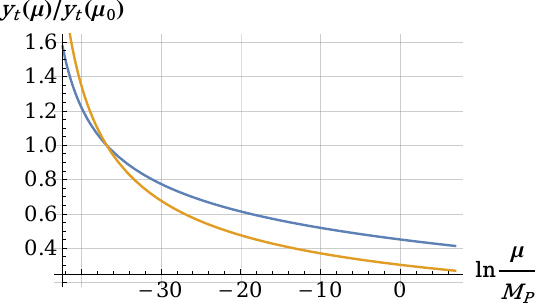}
    \end{subfigure} 
    \vspace{5pt}
    \\
     \begin{subfigure}[c]{0.48\textwidth}
         \centering
         \includegraphics[width=\textwidth]{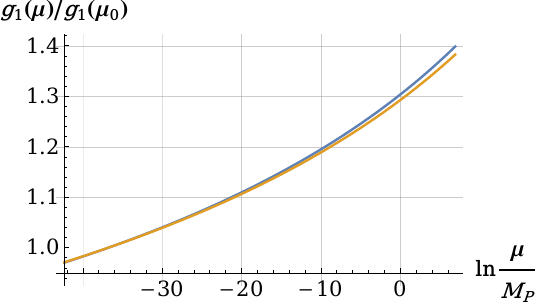}
     \end{subfigure}
     \hfill
     \begin{subfigure}[c]{0.48\textwidth}
         \centering
         \includegraphics[width=\textwidth]{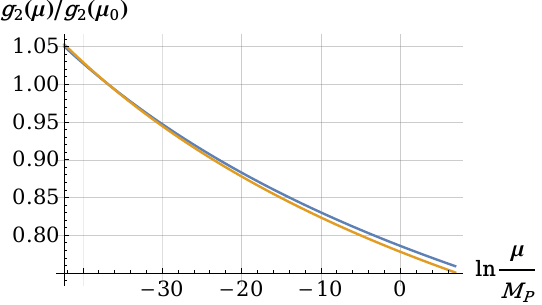}
    \end{subfigure}
    \vspace{15pt}
    \\
    \begin{subfigure}[c]{0.48\textwidth}
         \centering
         \includegraphics[width=\textwidth]{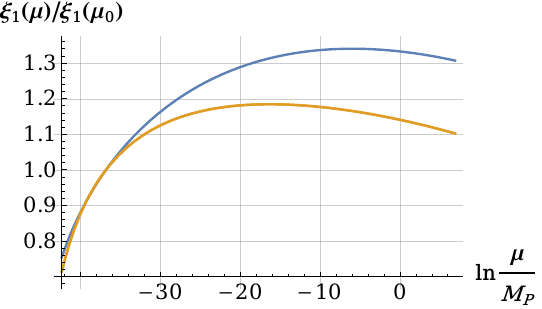}
     \end{subfigure}
     \hfill
     \begin{subfigure}[c]{0.48\textwidth}
         \centering
         \includegraphics[width=\textwidth]{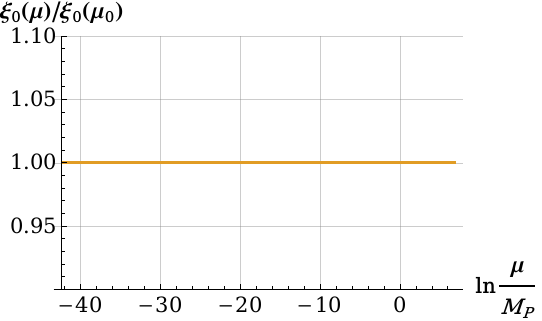}
     \end{subfigure}
    \caption{Running of the coupling constants $\lambda_0,\lambda_1,\lambda_2, \xi_0, \xi_1, g_1, g_2, g_3$ and $y_t$ with the energy scale $\mu$.}
    \label{run_plots_comp}
\end{figure}

As argued in Section \ref{Section:Quantum_Inflation}, we employ the constant subtraction scale choice, with $\mu=H_{inf}$, chosen by:
\begin{equation}
    H_{inf}^2 = \frac{V_1(\tau=10^2)}{3M_P^2},
\end{equation}
where $V_1(\tau)=V_{1-\textrm{loop}}(\tau,\mu=H_{inf})$. Figure \ref{run_plots_comp} illustrates the running of couplings with the energy scale $\mu$, where the initial conditions are given in equation (\ref{init_couplings}). All couplings' energy scale dependencies are initial conditions independent. Figure \ref{run_plots_comp} also depict the difference between regular RG run ($c_{\phi_i}= 1$) and with the suppression factors included $c_{\phi_i}\neq 1$. Figure \ref{tau_run_plt} illustrates the change in the value of the $\tau$ field with the inclusion of the $c_{\phi_i}$ factors and a comparison of their effects on $\tau$ running. In the numerical analysis of Section \ref{Section:Quantum_Inflation}, we neglect the running of $\lambda_0$, $\lambda_1$, and $\xi_0$, as the effects of changes in their values are negligible.

\newpage

\textcolor{white}{\\  
\\  
\\  
\\  
\\  
\\     }

\end{appendices}